\documentclass[11pt]{article}
\usepackage{graphicx, amssymb}
\newcommand{\SetFigFont}[3]{}

\title{An Integral Spectral Representation of the Propagator for
the Wave Equation in the Kerr Geometry}
\author{F.\ Finster, N.\ Kamran\thanks{Research supported by NSERC grant
\# RGPIN 105490-2004.}, J.\ Smoller\thanks{Research supported in
part by the NSF, Grant No.\ DMS-010-3998.}, and S.-T.\
Yau\thanks{Research supported in part by the NSF, Grant No.\
33-585-7510-2-30.}}
\date{June 2004}

\newtheorem{Def}{Def.}[section]
\newtheorem{Thm}[Def]{Theorem}
\newtheorem{Prp}[Def]{Proposition}
\newtheorem{Lemma}[Def]{Lemma}

\newtheorem{Corollary}[Def]{Corollary}
\newcommand{\Proof}{{\em{Proof. }}}
\newcommand{\QED}{\ \hfill $\FBox$ \\[1em]}
\newcommand{\spc}{\;\;\;\;\;\;\;\;\;\;}
\newcommand{\lbra}{\langle}
\newcommand{\lket}{\rangle}
\newcommand{\bra}{<\!\!}
\newcommand{\ket}{\!\!>}
\newcommand{\C}{\mathbb{C}}
\newcommand{\R}{\mathbb{R}}
\newcommand{\1}{\mbox{\rm 1 \hspace{-1.05 em} 1}}
\newcommand{\Z}{\mathbb{Z}}
\newcommand{\sZ}{\mbox{\rm \bf \scriptsize Z}}

\newcommand{\N}{\mathbb{N}}
\newcommand{\sN}{\mbox{\rm \scriptsize I \hspace{-.8 em} N}}

\newcommand{\beq}{\begin{equation}}
\newcommand{\eeq}{\end{equation}}

\newcommand{\FBox}{\rule{2mm}{2.25mm}}

\begin{document}

\maketitle
\begin{abstract}
We consider the scalar wave equation in the Kerr
geometry for Cauchy data which is smooth and compactly supported
outside the event horizon. We derive an integral representation
which expresses the solution as a superposition of solutions of
the radial and angular ODEs which arise in the separation of
variables. In particular, we prove completeness of the solutions
of the separated ODEs.

This integral representation is a suitable starting point for a detailed
analysis of the long-time dynamics of scalar waves in the Kerr geometry.
\end{abstract}
\tableofcontents

\newpage
\section{Introduction}
\setcounter{equation}{0}
In a recent paper~\cite{FKSY}, the long-term behavior of Dirac
spinor fields in the Kerr-Newman geometry,
which describes a charged rotating black hole in equilibrium, was investigated.
It was shown that solutions
of the Dirac equation for Cauchy data in $L^{2}$ outside
the event horizon and bounded near the event horizon, decay in
$L^{\infty}_{\mbox{\scriptsize{loc}}}$ as $t\to \infty$.
In this paper, we turn our attention to the scalar wave equation in the Kerr geometry. Our main result is to derive an integral representation
for the propagator, similar to the one obtained for the Dirac equation in~\cite{FKSY}.
In our next paper~\cite{W2}, we will use this
integral representation to analyze the long-time dynamics and
the decay of solutions in $L^\infty_{\mbox{\scriptsize{loc}}}$.

The analysis of the wave equation is quite different from that for the
Dirac equation. The main difficulty is that, in contrast to the Dirac equation,
there is no conserved density for the scalar wave equation which is
\emph{positive} everywhere outside the event horizon. This is due
to the fact that the charge density, which was positive for the Dirac
equation, is not positive for the wave equation. The other conserved
density, the energy density, is non-positive either: it is in general negative
inside the \emph{ergosphere}, a region outside the event horizon in which
the Killing vector corresponding to time translations becomes space-like.
For these reasons, it is not possible to introduce a positive scalar product
which is conserved in time. In more technical terms,
we are faced with the difficulty that it is impossible to represent the
Hamiltonian (i.e.\ the operator generating time translations)
as a selfadjoint operator on a Hilbert space.

We remark that the existence of the ergosphere is a direct
consequence of the fact that the Kerr black hole has angular
momentum~\cite{C}. Thus the ergosphere vanishes in the spherically
symmetric limit. This simplifies the analysis considerably.

A number of important contributions have been made to the rigorous study of
the scalar wave equation in black hole geometries. The current
last word on the stability of spherical black holes under scalar wave
perturbations is the paper by Kay and Wald~\cite{KW}, who proved using
energy estimates together with a reflection argument
that all solutions of the wave
equation in the Schwarzschild geometry are bounded in~$L^\infty$.
More recently, Klainerman, Machedon, and Stalker~\cite{SK} proved decay in
$L^\infty_{\mbox{\scriptsize{loc}}}$ of spherically symmetric solutions.
These papers use the spherical symmetry
of the Schwarzschild metric in an essential way. Whiting~\cite{W}
proved the absence of exponentially growing modes for the
Teukolsky equation with general spin $s=0, \frac{1}{2}, 1, \ldots$
(the case $s=0$ gives the scalar wave equation).
Whiting's approach is based on interesting differential and integral transforms,
which for a fixed angular momentum mode and fixed energy, convert the
reduced ODEs into an equation admitting a positive conserved energy.
Beyer~\cite{B} studied the wave and Klein-Gordon equations in
the Kerr metric, using an approach based on $C^{0}$ semigroup theory.
He proved that for each angular momentum mode, the Cauchy problem
is well-posed, and he also obtained a stability result for the
Klein-Gordon equation, provided that the mass
parameter in this equation is sufficiently large. Finally, Nicolas~\cite{Ni}
constructs a global solution for a non-linear Klein-Gordon equation
in Kerr.

Since the Hamiltonian cannot be represented as a selfadjoint operator on a
Hilbert space, we are forced to employ methods which are quite different
from those which we used in~\cite{FKSY}. More precisely,
the conserved energy gives rise to an {\em{indefinite}} scalar product, with
respect to which the Hamiltonian is selfadjoint. By considering the
system in finite volume with Dirichlet boundary conditions, we can arrange that
the scalar product is positive on the complement of a finite-dimensional
subspace. This allows us to use the general theory of Pontrjagin
spaces~\cite{Bo, L}. In particular, the Hamiltonian is essentially selfadjoint,
and has a spectral decomposition involving a finite set of complex
spectral points, which appear in complex conjugate pairs, together with a discrete
spectrum of real eigenvalues. We write the projectors onto the invariant subspaces
as contour integrals of the resolvent.
In order to obtain estimates for the resolvent, it is useful to consider the
Hamiltonian as a non-selfadjoint operator on a Hilbert space. This procedure
also works in the original infinite volume setting, and we derive operator
estimates which compare the resolvent in finite volume to that in infinite volume.
Using these estimates, we can represent the spectral projector corresponding to
the non-real spectrum as integrals over contours which are not closed and
lie inside a region of of the form $|{\mbox{Im}}\, \omega| < c\:
(1+|{\mbox{Re}}\, \omega|)^{-1}$ around the real axis.
At this point, we make use of the fact that the
scalar wave equation in the Kerr geometry is separable into ordinary differential
equations for the radial and angular parts~\cite{C}.
For the angular equation, we rely on the results of~\cite{FS}, where
a spectral representation is obtained for the angular
operator, and estimates for the eigenvalues and
spectral projectors are derived.
For the radial equation, we here derive
rigorous estimates which are based on the
semi-classical WKB approximation.
Using these estimates, we can express the resolvent in terms of solutions of the ODEs. Using furthermore Whiting's result that the ODEs admit no normalizable solutions for
complex $\omega$, we can deform the contours
onto the real line. This finally gives an integral representation for the propagator
in terms of the solutions of the ODEs with $\omega$ real.

To be more precise, recall that in Boyer-Lindquist
coordinates $(t, r, \vartheta, \varphi)$ with $r>0$, $0 \leq
\vartheta \leq \pi$, $0 \leq \varphi < 2\pi$, the Kerr metric
takes the form~\cite{C, HE}
\begin{eqnarray}
\lefteqn{ ds^2 \;=\; g_{jk}\:dx^j x^k } \nonumber \\ &=&
\frac{\Delta}{U} \:(dt \:-\: a \:\sin^2 \vartheta \:d\varphi)^2
\:-\: U \left( \frac{dr^2}{\Delta} + d\vartheta^2 \right) \:-\:
\frac{\sin^2 \vartheta}{U} \:(a \:dt \:-\: (r^2+a^2) \:d\varphi)^2
\spc \label{eq:0}
\end{eqnarray} with
\[ U(r, \vartheta) \;=\; r^2 + a^2 \:\cos^2 \vartheta \;,\spc
\Delta(r) \;=\; r^2 - 2 M r + a^2  \; , \] where $M$ and $aM$
denote the mass and the angular momentum of the black hole,
respectively. We shall restrict attention to the case $M^2 \geq a^2$,
because otherwise there is a naked singularity.
In the {\em{non-extreme case}} $M^2 > a^2$, the
function $\Delta$ has two distinct zeros,
\[ r_0 \;=\; M \:-\: \sqrt{M^2 - a^2} \spc {\mbox{and}} \spc r_1 \;=\; M \:+\: \sqrt{M^2 -
a^2} \; , \] corresponding to the Cauchy and the event horizon,
respectively. In the {\em{extreme case}} $M^2=a^2$, the
Cauchy and event horizons coincide,
\[ r_0 \;=\; r_1 \;=\; M\:. \]
We shall consider only the region $r>r_1$
outside the event horizon, and thus $\Delta>0$.

In order to determine the ergosphere, we consider the norm of
the Killing vector $\xi=\frac{\partial}{\partial t}$,
\begin{equation}
g_{ij}\,\xi^{i}\xi^{j} \;=\; g_{tt} \;=\; \frac{\Delta-a^{2}\,\sin^{2}\vartheta}{U}
\;=\; \frac{r^2 - 2Mr + a^2 \cos^2 \vartheta}{U}\: .
\end{equation}
This shows that $\xi$ is space-like in the open region of
space-time where
\begin{equation}
r^{2}-2Mr+a^{2}\,\cos^{2}\vartheta \;<\; 0 \:,
\end{equation}
the so-called {\em{ergosphere}}. It is a bounded region of space
outside the event horizon, and intersects the event horizon at the poles
$\vartheta=0,\,\pi$.

The scalar wave equation in the Kerr geometry is
\begin{equation} \label{swave}
\square \,\Phi:= g^{ij}\nabla_{i}\nabla_{j}\,\Phi=\frac{1}{\sqrt
{-g}}\,\frac{\partial}{\partial
x^{i}} \left( {\sqrt{-g}}\,g^{ij}\frac{\partial}{\partial
x^{j}} \right) \Phi \;=\; 0 \:,
\end{equation}
where $g$ denotes the determinant of the metric $g_{ij}$.
In Boyer-Lindquist coordinates this becomes
\begin{eqnarray} \label{wave}
\lefteqn{ \left[ -\frac{\partial}{\partial r}\Delta\frac{\partial}{\partial r}
+\frac{1}{\Delta} \left( (r^{2}+a^{2})\frac{\partial}{\partial
t}+a\frac{\partial}{\partial \varphi} \right)^{2}
-\frac{\partial}{\partial \cos \vartheta} \sin^{2}\vartheta
\frac{\partial}{\partial \cos \vartheta} \right. \nonumber } \\
&& \hspace*{5.5cm} \left. -\frac{1}{\sin^{2}\vartheta} \left( a\sin^{2}\vartheta
\frac{\partial}{\partial t}+\frac{\partial}{\partial
\varphi}\right)^{2} \right ]\Phi \;=\; 0\:.
\end{eqnarray}
In what follows, we denote the square bracket in this equation by $\square$ (although
strictly speaking, it is a scalar function times the wave operator
in~(\ref{swave})).

A key property of the wave equation in the Kerr metric is that can be separated into ordinary
differential equations by making the usual multiplicative ansatz
\begin{equation}\label{separansatz}
\Phi(t,r,\vartheta,\varphi)=e^{-i\omega
t-ik\varphi}\:R(r)\:\Theta(\vartheta),
\end{equation}
where $\omega$ is a quantum number which could be real or complex
and which corresponds to the ``energy'', and $k$ is an integer
quantum number corresponding to the projection
of angular momentum onto the axis of symmetry of the black
hole. Substituting~(\ref{separansatz}) into~(\ref{wave}), we see that
\begin{equation} \label{wavesep}
\square \Phi \;=\; ({\cal{R}}_{\omega , k}+{\cal{A}}_{\omega , k})\Phi,
\end{equation}
where ${\cal{R}}_{\omega , k}$ and ${\cal{A}}_{\omega , k}$ are
the radial and angular operators
\begin{eqnarray} \label{radial}
{\cal{R}}_{\omega , k} &=& \, -\frac{\partial}{\partial r}
\Delta \frac{\partial}{\partial r}
- \frac{1}{\Delta}((r^{2}+a^{2})\omega +ak)^{2} \\
{\cal{A}}_{\omega , k} &=& \label{angular}
-\frac{\partial}{\partial \cos \vartheta}\: \sin^2 \vartheta\:
\frac{\partial}{\partial \cos \vartheta}
+\frac{1}{\sin^2 \vartheta}(a\omega \sin^2 \vartheta + k )^{2}.
\end{eqnarray}
We can therefore separate the variables $r$ and $\vartheta$ to obtain for
fixed $\omega$ and $k$ the system of ODEs
\begin{equation} \label{odes}
{\cal{R}}_{\omega ,k}\,R_\lambda \;=\; -\lambda \, R_{\lambda},\qquad  {\cal{A}}_{\omega ,k}\,\Theta_\lambda \;=\; \lambda \, \Theta_\lambda\:,
\end{equation}
where the separation constant $\lambda$ is an eigenvalue of the
angular operator ${\cal{A}}_{\omega, k}$ and can thus be regarded as
an angular quantum number. In the spherically symmetric case (i.e.\ $a=0$),
$\lambda$ goes over to the usual eigenvalues $\lambda=l(l+1)$ of
total angular momentum.
Since the $k$-modes are obtained simply by expanding the $\varphi$-dependence in a Fourier series, we can in what follows restrict attention to one fixed $k$-mode and
omit the index $k$. We point out that for the $\lambda$-modes the situation is more difficult because $\lambda$ as well as the corresponding angular eigenfunction $\Theta_\lambda(\vartheta)$ will in general depend on $\omega$. As a consequence, it is at this point not clear how to decompose the initial data into $\lambda$-modes.

We now reformulate the wave equation in first-order Hamiltonian
form. The resolvent of this Hamiltonian will be one of the main
ingredients in the statement of our main theorem. Letting
\begin{equation}
\Psi=\left(
\begin{array}{c} \Phi \\
i\partial_{t}\Phi
\end{array} \right),
\end{equation}
the wave equation (\ref{wave}) takes the form
\begin{equation}\label{Hamform}
i\,\partial_{t}\Psi=H\,\Psi,
\end{equation}
where $H$ is the Hamiltonian
\begin{equation} \label{hamil}
H=\left(
\begin{array}{cc}
0 & 1 \\
\alpha & \beta
\end{array}
\right).
\end{equation}
Here $\alpha$ and $\beta$ are the differential operators
\begin{eqnarray*}
\alpha &=& \left(\frac{(r^{2}+a^{2})^{2}} {\Delta}-a^{2}\sin^{2}\vartheta\right)^{-1}
\left[-{\partial}_{r}\Delta\,\partial_{r} -\partial_{\cos \vartheta}\sin^{2}\vartheta \partial_{\cos
\theta}\, +\left(\frac{a^2}{\Delta}-\frac{1}{\sin^{2}\vartheta} \right)
\partial^{2}_{\varphi}\right] \\
\beta &=& -2a \left(\frac{(r^{2}+a^{2})^{2}}{\Delta}
-a^{2}\sin^{2}\vartheta\right)^{-1}\left(\frac{r^{2}+a^{2}}{\Delta}-1\right) i\partial_{\varphi} \:.
\end{eqnarray*}

\begin{figure}[tbp]
\begin{center}
\input{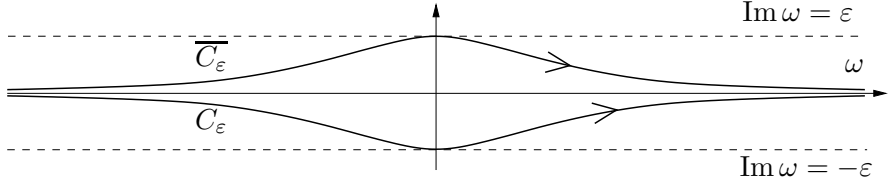}
\caption{Integration contours}
\label{fig0}
\end{center}
\end{figure}

We now state our main result.
\begin{Thm} \label{thm1}
Consider the Cauchy problem
\begin{equation}
\square\,\Phi \;=\;0 \:,\spc
(\Phi, i \partial_{t}\Phi)(0,x) \;=\; \Psi_0(x)
\end{equation}
for initial data $\Psi_0 \in C^\infty_0((r_1, \infty) \times S^2)^2$
which is smooth and compactly supported outside the event horizon, in the
slicing associated to the Boyer Lindquist coordinates.
Then there exists a unique global
solution $\Psi(t) = (\Phi(t), i \partial_t \Phi(t))$ which can be
represented as follows,
\begin{eqnarray}
\lefteqn{ \Psi(t,r,\vartheta, \varphi) } \nonumber \\
&=& -\frac{1}{2\pi i} \sum_{k \in \sZ} e^{-i k \varphi} \sum_{n \in \sN}\;
\lim_{\varepsilon \searrow 0} \left( \int_{C_\varepsilon} -\int_{\overline{C_\varepsilon}} \right) d\omega\; e^{-i \omega t}\:
(Q_{k,n}(\omega)\: S_{\infty}(\omega)\: \Psi_0^k)(r, \vartheta) \:.\quad \label{intrep}
\end{eqnarray}
Here the sums and integrals converge in $L^2_{\mbox{\scriptsize{loc}}}$.
\end{Thm}
In the statement of this theorem we use the following notation.
The function~$\Psi_0^k$ is the $k$th angular Fourier component of $\Psi_0$, i.e.
\[ \Psi_0^k(r, \vartheta) \;=\; \frac{1}{2 \pi}  \int_0^{2 \pi} e^{i k \varphi}\: \Psi_0(r,\vartheta, \varphi)\: d\varphi\:. \]
We consider $\omega$ in the lower complex half plane $\{ {\mbox{Im}}\,
\omega <0 \}$, and
$C_\varepsilon$ is a contour which joins the points $\omega=-\infty$ with
$\omega=\infty$ and stays in an
$\varepsilon$-neighborhood of the real line. A typical example is
\[ C_\varepsilon \;=\; \{ x-i \varepsilon \:e^{-x^2} \::\: x \in \R\} . \]
$\overline{C_\varepsilon}$ is the complex conjugated contour. Thus the integrals in~(\ref{intrep}) can be thought of as a ``contour integral
around the real axis'' (see Figure~\ref{fig0}), in analogy to the
well-known Cauchy integral formula for matrices
\[ e^{-i A t} \;=\; -\frac{1}{2 \pi i} \oint_C e^{-i \omega t}\:
(A-\omega)^{-1}\: d\omega \;, \]
where $A$ is a finite-dimensional matrix and $C$ a contour which
encloses the whole spectrum of $A$.
For given $\omega$ and $k$, the wave operator is a sum of a radial operator
${\cal{R}}_{\omega, k}$ and an angular operator ${\cal{A}}_{\omega, k}$.
As shown in~\cite{FS}, the angular operator has for $\omega$ near the real line
a purely discrete spectrum consisting of
eigenvalues $(\lambda_n)_{n \in \sN}$ (see Lemma~\ref{lemmaangular}).
The spectral projectors onto the
corresponding eigenspaces, which are one-dimensional, are denoted by
$Q_{k,n}(\omega)$. Furthermore, we write the wave equation
in the Hamiltonian form~(\ref{Hamform}, \ref{hamil}) and let
$S_\infty(\omega)=(H-\omega)^{-1}$ be the resolvent (in a suitable Sobolev space).
The operator product $Q_{k,n}\:S_\infty$ can be expressed in terms of
solutions of the reduced ordinary differential equations
(see Proposition~\ref{thm63} with $g=s$ and $s$ as in Lemma~\ref{lemma61}).
Relying on Whiting's mode stability~\cite{W}, we shall see below
that the integrand in~(\ref{intrep}) is well-defined and
holomorphic in the lower half plane, and thus the value of the integrals is indeed independent of the choice of $C_\varepsilon$.
If the integrand were continuous up to the real axis, we could
in~(\ref{intrep}) take the limit $\varepsilon \searrow 0$
to obtain an integral along the real line. However, we do not know
whether the integrand in~(\ref{intrep}) is continuous on
the real axis. Thus the integral in~(\ref{intrep}) can be regarded
as an integral over the real axis, with an ``$i \varepsilon$-regularization
procedure'' of possible singularities (if for instance the integrand
had a simple pole at $\omega_0 \in \R$, this would give
rise to a $\delta$-contribution at $\omega_0$).

We point out that the global existence and uniqueness of solutions of
the Cauchy problem can be obtained more generally in globally hyperbolic
space-times (see e.g.~\cite{Le}).
The main point of Theorem~\ref{thm1} is that we give an explicit
decomposition of the propagator as a superposition of solutions of the ODEs
which arise in the separation of variables. In particular, Theorem~\ref{thm1}
shows completeness in the sense that the solutions of the coupled ODEs
for real~$\omega$ form a basis of the solution space.
The explicit form of~(\ref{intrep}) is useful for the study of the dynamics,
because the time-dependence of $\Psi$ is related by a simple Fourier
transform to the $\omega$-dependence of the integrand in~(\ref{intrep}), which can in turn be analyzed by getting suitable ODE estimates~\cite{W2}.

We finally remark that the case of the wave equation for a
scalar field minimally coupled to an electromagnetic field,
\begin{equation}
g^{jk}(\nabla_{j}-ieA_{j})(\nabla_{k}-ieA_{k})\,\Phi=0,
\end{equation}
could be treated by similar methods in the non-extreme Kerr-Newman
geometry, where now the metric is given by (\ref{eq:0}) with
\begin{equation}
U(r, \vartheta) \;=\; r^2 + a^2 \:\cos^2 \vartheta \;,\spc
\Delta(r) \;=\; r^2 - 2 M r + a^2 + q^2 \; ,
\end{equation}
and the electromagnetic potential is
\begin{equation}
A_j \:dx^j \;=\; -\frac{q \:r}{U} \:(dt \:-\:  a \:\sin^2
\vartheta \: d\varphi) \; ,
\end{equation}
where $q$ denotes the charge of the black hole, and the parameters
$M,a,q$ satisfy the inequality $M^{2}>a^{2}+q^{2}$.

\section{Preliminaries}
\setcounter{equation}{0}
In this section we briefly recall the variational formulation of the
wave equation and the separation of variables. Furthermore, we
bring the equation into a first-order Hamiltonian form. Finally, we
introduce and discuss scalar products which
are needed for the construction of the propagator.

The wave equation~(\ref{wave}) is the Euler-Lagrange equation corresponding to the action
\begin{equation}
S \;=\; {\int_{-\infty}^{\infty}}\, dt
\int_{r_1}^{\infty}dr\int_{-1}^{1}d(\cos
\vartheta)\int_{0}^{\pi}d\varphi\,{\cal{L}}(\Phi,\nabla \Phi) \;,
\end{equation}
where the Lagrangian ${\cal{L}}$ is given by
\begin{eqnarray}
{\cal{L}}&=&-\Delta|\partial_{r}\Phi|^{2}+\frac{1}{\Delta} \left|((r^{2}+a^{2})\partial_{t}
+a\partial_{\varphi})\Phi \right|^{2} \nonumber \\
&& -\sin^{2}{\vartheta} \left|\partial_{\cos \vartheta}\varphi \right|^{2} -\frac{1}{\sin^{2}\vartheta} \left| (a\sin^{2}\vartheta\partial_{t}+\partial_{\varphi})\Phi \right|^{2}. \label{lagr}
\end{eqnarray}
According to Noether's theorem, symmetries of the Lagrangian give rise to conserved quantities.
The symmetry under local gauge transformations yields that the vector field
\begin{equation}
J_{k} \;=\; -{\mbox{Im}}\,({\overline{\Phi}}\: \nabla_{k}\Phi)\:,
\end{equation}
called the (electromagnetic) current,
is divergence free, and integrating the normal component of this current
over the hypersurface $t={\mbox{const}}$ yields the conserved charge
$Q$. More precisely,
\[ Q[\Phi] \;=\; \int_{r_1}^{\infty} dr \int_{-1}^{1}d(\cos \vartheta)
\int_{0}^{2\pi}\frac{d\varphi}{2 \pi} \:{\cal{Q}}\,, \]
where ${\cal{Q}}$ is the charge density
\begin{eqnarray*}
{\cal{Q}} &=& i\frac{\partial {\cal L}}{\partial \Phi_{t}}\,\Phi \\
&=& {\mbox{Re}} \left\{
\frac{(r^{2}+a^{2})^{2}}{\Delta}\:
\overline{\Phi} \left(i\partial_{t}\Phi+\frac{a\:i\partial_\varphi \Phi}{r^{2}+a^{2}} \right)
-a^{2}\sin^{2}\vartheta \:\overline{\Phi}
\left(i \partial_{t}\Phi+\frac{i \partial_{\varphi}\Phi}{a
\sin^{2}\vartheta}\right) \right\} .
\end{eqnarray*}
Moreover, since the Kerr metric is stationary, the Lagrangian
is invariant under time translations. The corresponding conserved
quantity is the energy $E$,
\begin{equation}\label{energy}
E[\Phi] \;=\; \int_{r_1}^{\infty} dr \int_{-1}^{1} d(\cos\vartheta)
\int_{0}^{2\pi} \frac{d\varphi}{2 \pi} \:\cal{E},
\end{equation}
where ${\cal{E}}$ is the energy density
\begin{eqnarray}
{\cal{E}} \;=\; \frac{\partial \cal{L}}{\partial
\Phi_{t}}\,\Phi_{t}-{\cal{L}} &=&
\left({\frac{(r^{2}+a^{2})^{2}}{\Delta}} - a^{2}\,\sin^{2}\vartheta \right) \left|\partial_{t} \Phi \right|^{2}+\Delta
\left|\partial_{r}\Phi \right|^{2} \nonumber \\
&&+\sin^{2}\vartheta \left|\partial_{\cos
\vartheta}\Phi \right|^{2}+\left(
{\frac{1}{\sin^{2}\vartheta}}-{\frac{a^{2}}{\Delta}}\right) \left| \partial_{\varphi}\Phi \right|^{2} .
\label{energydensity}
\end{eqnarray}
One sees that all the terms in the energy density are positive, except for
the coefficient of $|\partial_{\varphi}\Phi|^{2}$, which is
positive if and only if $r^{2}-2Mr+a^{2}\,\cos^{2}\vartheta >0$, i.e.\
outside the ergosphere. As a consequence, $E$ is in general not positive.

Our analysis is based on a few properties of the angular operator ${\cal{A}}_\omega$, which we now
state. For real $\omega$, the angular operator ${\cal{A}}_\omega$
clearly is formally selfadjoint on $L^2(S^2)$. However, this is not
sufficient for our purpose, because we need to consider
the case that $\omega$ is complex.
In this case, ${\cal{A}}_\omega$ is a non-selfadjoint operator.
Nevertheless, we have the following spectral decomposition, which is
proved in~\cite{FS}.
\begin{Lemma} {\bf{(angular spectral decomposition)}} \label{lemmaangular}
For any given $c>0$, we define the open set $U \subset \C$ by the condition
\beq \label{ocond}
|{\mbox{\rm{Im}}}\, \omega| \;<\; \frac{c}{1 + |{\mbox{\rm{Re}}}\, \omega|}\:.
\eeq
Then there is an integer~$N$ and a
family of operators $Q_n(\omega)$ defined for
$n \in \N \cup \{0\}$ and $\omega \in U$
with the following properties:
\begin{description}
\item[(i)]
The $Q_n$ are holomorphic in~$\omega$.
\item[(ii)]
$Q_0$ is a projector onto an $N$-dimensional invariant
subspace of ${\cal{A}}_\omega$. For $n>0$, the $Q_n$ are projectors
onto one-dimensional eigenspaces of ${\cal{A}}_\omega$ with corresponding
eigenvalues $\lambda_n(\omega)$. These eigenvalues satisfy a bound of
the form
\beq \label{ange1}
|\lambda_n(\omega)| \;\leq\; C(n)\: (1+|\omega|)
\eeq
for suitable constants $C(n)$. Furthermore, there is a parameter
$\varepsilon>0$ such that for all $n \in \N$ and $\omega \in U$,
\beq \label{ang11}
|\lambda_n(\omega)| \;\geq\; n\: \varepsilon\:.
\eeq
\item[(iii)] The $Q_n$ are complete, i.e.\
\[ \sum_{n=0}^\infty Q_n \;=\; \1 \]
with strong convergence of the series in~$L^2(S^2)$.
\item[(iv)] The $Q_n$ are uniformly bounded in $L^2(S^2)$,
i.e. for all $n \in \N_0$,
\beq \label{Qnb}
\|Q_n\| \;\leq\; c_1
\eeq
with $c_1$ independent of $\omega$ and $n$.
\end{description}

If $c$ is sufficiently small, $c<\varepsilon$, or
the real part of $\omega$ is sufficiently large,
$|{\mbox{\em{Re}}}\, \omega| > C(c)$, one can choose $N=1$, i.e.\
${\cal{A}}_\omega$ is diagonalizable with non-degenerate eigenvalues.
\end{Lemma}
The proof of this lemma is outlined as follows. If $|{\mbox{Im}}\, \omega|$
is sufficiently small, the imaginary part of the potential
can be treated as a slightly non-selfadjoint perturbation
(see~\cite[Chapter~5, {\S} 4.5]{Kato}), giving a spectral decomposition
into one-dimensional eigenspaces. On the other hand, for any fixed
$\omega \in \C$, an asymptotic analysis of the resolvent $({\cal{A}}_\omega
-\lambda)^{-1}$ for large $\lambda$ (see e.g.~\cite[Chapter~12]{CL}) yields
a spectral decomposition into invariant subspaces which for large $|\lambda|$
are one-dimensional eigenspaces. Thus the
difficult point in the above lemma is to show that $N$ and the
constant $c_1$ can be chosen uniformly in $\omega \in U$. To this end, one
must show that for real $\omega$, the eigenvalue gaps of the selfadjoint
operator ${\cal{A}}_\omega$ become large as $|\omega| \to \infty$.
These gap estimates are worked out in~\cite{FS} by analyzing the solutions
of the corresponding complex Riccati equation.

After separation~(\ref{separansatz}), the
reduced wave equation takes the form
\begin{eqnarray}
\lefteqn{ \left[ -\frac{\partial}{\partial r}\Delta\frac{\partial}{\partial r}
-\frac{1}{\Delta} \left( (r^{2}+a^{2}) \omega+ak \right)^{2} \right.
\nonumber } \\
&& \left. -\frac{\partial}{\partial \cos \vartheta} \sin^{2}\vartheta
\frac{\partial}{\partial \cos \vartheta}
+\frac{1}{\sin^{2}\vartheta}(a\omega \sin^{2}\vartheta + k
)^{2} \right] \Phi \;=\; 0. \label{wred}
\end{eqnarray}
Under the separation, the above expressions for the charge and energy
densities become
\begin{eqnarray}
{\cal{Q}} &=& |\Phi|^{2} \left\{ \frac{(r^{2}+a^{2})^{2}}{\Delta}
\left( {\mbox{Re}}\,\omega + \frac{ak}{r^{2}+a^{2}} \right)-a^{2}\sin^{2}\vartheta \left( {\mbox{Re}}\,
\omega +\frac{k}{a \sin^{2}\vartheta} \right) \right\}
\label{qsep} \\
{\cal{E}} &=& |\Phi|^{2} \left\{ \frac{(r^{2}+a^{2})^{2}}{\Delta}
\left( |\omega|^2 - \frac{a^2 k^2}{(r^{2}+a^{2})^2} \right)-a^{2}\sin^{2}\vartheta \left( |\omega|^2 -\frac{k^2}{a^2 \sin^{4}\vartheta} \right) \right\} \nonumber \\
&&+\Delta \left|\partial_{r}\Phi \right|^{2} + \sin^{2}\vartheta \left| \partial_{\cos \vartheta}\Phi \right|^{2} . \label{esep}
\end{eqnarray}

It is a subtle point to find a scalar product $\bra .,. \ket$
which is well-suited to the analysis of the wave equation.
It is desirable to choose the scalar product such that the Hamiltonian
$H$ is Hermitian (i.e.\ formally selfadjoint) with respect to it. Since $H$
is the infinitesimal generator of time translations, $H$ is Hermitian
w.r.\ to $\bra .,. \ket$ if and only if the inner product
$\bra \Psi, \Psi \ket$ is time independent for all solutions $\Psi=(\Phi, i \partial_t \Phi)$ of
the wave equation. This can for example be achieved by imposing that
$\bra \Psi, \Psi \ket$ should be equal to the energy $E$ corresponding to $\Psi$.
This leads us to introduce a scalar product by polarizing the formula
for the energy, (\ref{energy}, \ref{energydensity}). We thus obtain
the so-called {\em{energy scalar product}}
\begin{eqnarray}
\lefteqn{ \bra \Psi, \Psi'\ket \;=\; \int_{r_1}^{\infty}dr\int_{-1}^{1} d(\cos
\vartheta)  \left\{
\left({\frac{(r^{2}+a^{2})^{2}}{\Delta}} - a^{2}\,\sin^{2}\vartheta \right) \overline{\partial_{t} \Phi}\, \partial_{t}{\Phi}' \right. \nonumber } \\
&& \left. +\Delta \, \overline{\partial_{r}\Phi}
\,\partial_{r}{\Phi}' +\sin^{2}\vartheta
\,\overline{\partial_{\cos \vartheta}\Phi}\,\partial_{\cos
\vartheta}{\Phi}'  +\left(
{\frac{1}{\sin^{2}\vartheta}}-{\frac{a^{2}}{\Delta}}\right)\,\overline{\partial_{\varphi}\Phi}
\,\partial_{\varphi}{\Phi}' \right\}, \label{energysp}
\end{eqnarray}
where again $\Psi = (\Phi, i \partial_t \Phi)$ and $\Psi' = (\Phi', i \partial_t \Phi')$.
If $\Psi'$ is a solution of the reduced system of ODEs~(\ref{odes}),
$\Psi'$ can be written as $\Psi' = (\Phi_{\omega, \lambda}, \omega \Phi_{\omega, \lambda})$ with $\Phi_{\omega, \lambda}(r, \vartheta) = R_\lambda(r) \Theta_\lambda(\vartheta)$.
Integrating by parts and dropping the boundary terms (which is certainly
admissible when we consider the system in finite volume or when
$\Psi$ has compact support), we can substitute
the radial and angular equations into~(\ref{energysp}) to obtain
\begin{eqnarray}
\lefteqn{ \bra \Psi, \Psi_{\omega,\lambda} \ket \;=\; \omega \int_{r_1}^\infty dr \int_{-1}^1 d (cos\vartheta) } \nonumber \\
&&\times \left[
\left( \frac{(r^2+a^2)^2}{\Delta} - a^2 \sin^2 \vartheta \right) \overline{(i \partial_t + \overline{\omega}) \Phi}
\:\Phi_{\omega,\lambda} + 2 a k \left(\frac{r^2+a^2}{\Delta}-1 \right) \overline{\Phi} \:\Phi_{\omega, \lambda} \right]. \quad
\label{5jprod}
\end{eqnarray}
In the special case $\Psi=\Psi'$, this reduces to
\begin{eqnarray}
\lefteqn{ \bra \Psi_{\omega,\lambda}, \Psi_{\omega,\lambda} \ket
\;=\; 2 \omega \int_{r_1}^\infty dr \int_{-1}^1 d (cos\vartheta)
\: |\Phi_{\omega, \lambda}|^2 } \nonumber \\
&&\times \left[ \left( \frac{(r^2+a^2)^2}{\Delta} - a^2 \sin^2 \vartheta \right) {\mbox{Re}}\, \omega \:+\: a k\left(\frac{r^2+a^2}{\Delta}-1 \right) \right]. \label{5iprod}
\end{eqnarray}
By construction, the Hamiltonian is Hermitian with respect to the
energy scalar product. However, the energy scalar product is in general
not positive definite. This is obvious in~(\ref{energysp}) because
the factor $(\sin^{-2} \vartheta - a^2/\Delta)$ is negative inside
the ergosphere. Likewise, the integrand in~(\ref{5iprod})
can be negative because the factor $ak$ in the
second term in the brackets can have any sign.

Apart from the energy, also the charge~$Q$ gives rise to a conserved scalar product.
It is a natural idea to try to obtain a positive scalar product by taking a suitable linear combination of
these two scalar products. Unfortunately, comparing~(\ref{esep}) and~(\ref{qsep})
one sees that it is impossible to form a non-trivial linear combination of
${\cal{Q}}$ and ${\cal{E}}$ which is manifestly positive everywhere.
One might argue that a suitable linear combination might nevertheless
be positive because the positive term $\Delta |\partial_r \Phi|^2 +
\sin^2 \vartheta |\partial_{\cos \vartheta} \Phi|^2$ might compensate
the negative terms. However, comparing~(\ref{5iprod}) with~(\ref{qsep}), one sees
that there is a simple relation between the energy scalar product and the charge,
\[ \bra \Psi_{\omega,\lambda}, \Psi_{\omega,\lambda} \ket \;=\;
2 \omega\: Q[\Psi_{\omega,\lambda}] \;, \]
making it again impossible to form a linear combination such that the
integrand of the corresponding scalar product is everywhere positive.
Stephen Anco showed that it is indeed impossible to introduce
a conserved density for the wave equation
which gives rise to a positive definite scalar product~\cite{A}.
We conclude that if we
want to consider $H$ as a selfadjoint operator, the underlying scalar
product will necessarily be indefinite.

But we can clearly consider $H$ as a non-selfadjoint operator on a Hilbert space,
and this point of view will indeed be useful for the estimates of Section~\ref{sec5}.
Our method for constructing a positive scalar product is to simply replace
the negative term $-a^2/\Delta$ in~(\ref{energysp}) by a positive term.
More precisely, we introduce the scalar product $(.,.)$ by
\begin{eqnarray}
\lefteqn{ ( \Psi, \Psi' ) \;=\; \int_{r_1}^{\infty}dr\int_{-1}^{1} d(\cos
\vartheta) \left\{
\left({\frac{(r^{2}+a^{2})^{2}}{\Delta}} - a^{2}\,\sin^{2}\vartheta \right) \overline{\partial_{t} \Phi}\, \partial_{t}{\Phi}'
\right. \nonumber } \\
&& \left. +\Delta \, \overline{\partial_{r}\Phi}
\,\partial_{r}{\Phi}' +\sin^{2}\vartheta
\,\overline{\partial_{\cos \vartheta}\Phi}\,\partial_{\cos
\vartheta}{\Phi}'  +
{\frac{1}{\sin^{2}\vartheta}} \,\overline{\partial_{\varphi}\Phi}
\,\partial_{\varphi}{\Phi}'
+ \frac{(r^2+a^2)^2}{\Delta}\:\overline{\Phi} \,{\Phi'}
\right\} . \qquad\label{psp}
\end{eqnarray}
We denote the corresponding Hilbert space by ${\cal{H}}$ and the norm by $\|.\|$.
This norm dominates the energy scalar product
in the sense that there is a constant $c_1>0$ depending only on the
geometry such that the ``Schwarz-type'' inequality
\begin{equation} \label{schwarz}
|\bra \Psi, \Psi' \ket| \;\leq\; c_1\: \|\Psi\|\: \|\Psi'\|
\end{equation}
holds for all $\Psi, \Psi' \in {\cal{H}}$.

We finally bring the Hamiltonian and the above inner products into a
more convenient form. First, we introduce the Regge-Wheeler
variable~$u$ by
\begin{equation} \label{51a}
\frac{du}{dr} \;=\; \frac{r^2+a^2}{\Delta} \;,\spc
\frac{\partial}{\partial r} \;=\; \frac{r^2+a^2}{\Delta}\: \frac{\partial}{\partial u}
\:.
\end{equation}
The variable $u$ ranges over $(-\infty, \infty)$ as $r$ ranges over $(r_1, \infty)$.
Furthermore, we introduce the functions
\begin{eqnarray}
\rho &=& r^2+a^2 \:-\: a^2\: \sin^2 \vartheta\: \frac{\Delta}{r^2+a^2} \label{rdef} \\
\beta &=& -\frac{2 a k}{\rho} \left( 1 - \frac{\Delta}{r^2+a^2} \right) \\
\delta &=& \frac{1}{\rho} \left( r^2+a^2\:+\: \frac{a^2 k^2}{r^2+a^2} \right) \label{ddef}
\end{eqnarray}
as well as the operator
\beq \label{Adef}
A \;=\; \frac{1}{\rho} \left[-\frac{\partial}{\partial u} (r^2+a^2)
\frac{\partial}{\partial u} - \frac{\Delta}{r^2+a^2}\: \Delta_{S^2}
- \frac{a^2 k^2}{r^2+a^2} \right] ,
\eeq
where $\Delta_{S^2}$ denotes the Laplacian on the 2-sphere (recall that the
parameter $k$ is fixed throughout). Then, after integrating by parts,
our inner products can on $C^2(\R \times S^2)^2$ be written as
\begin{eqnarray}
\bra \Psi_1, \Psi_2 \ket &=& \int_{-\infty}^\infty du \int_{-1}^1 d\cos \vartheta
\: \rho\; \lbra \Psi_1, \left( \begin{array}{cc} A & 0 \\ 0 & 1 \end{array} \right)
\Psi_2 \lket_{\C^2} \label{ESP} \\
( \Psi_1, \Psi_2 ) &=& \int_{-\infty}^\infty du \int_{-1}^1 d\cos \vartheta
\: \rho\; \lbra \Psi_1, \left( \begin{array}{cc} A+\delta & 0 \\ 0 & 1 \end{array}
\right) \Psi_2 \lket_{\C^2} \:, \label{PSP}
\end{eqnarray}
and the Hamiltonian takes the form
\beq \label{HAb}
H \;=\; \left( \begin{array}{cc} 0 & 1 \\ A & \beta \end{array}
\right) .
\eeq
The functions $\rho$, $\beta$, and $\delta$ satisfy for a suitable constant
$c>0$ the bounds
\[ \frac{1}{c} \;\leq\; \frac{\rho}{r^2+a^2} \;\leq\; c \;,\qquad
|\beta|, |\delta| \;\leq\; c \:. \]
We abbreviate the integration measure in~(\ref{ESP}) and~(\ref{PSP}) by
\beq \label{mudef}
d\mu \;:=\; \rho\: du\: d\cos \vartheta\:.
\eeq

\section{Spectral Properties of the Hamiltonian in a Finite Box} \label{sec3}
\setcounter{equation}{0} We saw in the preceding section that the
energy inner product (\ref{energysp}), with respect to which the
Hamiltonian is formally selfadjoint,
is in general indefinite. This fact remains
true even when as in~\cite{FKSY} we consider the system in a
``finite box,'' i.e.\ when the range of the radial variable $r$
is restricted to a bounded interval $r \in [r_L, r_R]$ with
$r_1 < r_L < r_R < \infty$.
Accordingly, in order to derive a spectral representation for the
propagator corresponding to the wave equation (\ref{wave}),
we will need to consider the spectral theory of operators on
indefinite inner product spaces.
Since there is an extensive literature on this topic, we here only
recall the basic facts needed for our analysis, referring the
reader to \cite{Bo, L} for details.

A \emph{Krein space} is a complex vector space
${\cal{K}}$ endowed with a non-degenerate inner product $\bra .\:,\:
.\ket$ and an orthogonal direct sum decomposition
\begin{equation}
{\cal{K}}={\cal{K}}_{+}\oplus {\cal{K}}_{-},
\end{equation}
such that $({\cal{K}}_{+}, \bra .\:,\: .\ket)$ and
$({\cal{K}}_{-}, -\bra .\:,\: .\ket)$ are both Hilbert spaces.
A selfadjoint operator $A$ on a Krein space ${\cal{K}}$ is said to
be \emph{definitizable} if there exists a non-constant real polynomial
$p$ of degree~$k$ such that
\begin{equation}
\bra p\,(A)\,x\:,\:x \ket \;\geq\; 0
\end{equation}
for all $x\in {\cal{D}}(A^{k})$. Definitizable
operators have a spectral decomposition, which is similar to the
spectral theorem in Hilbert spaces, except that there is in general an
additional finite point spectrum in the complex plane (see~\cite[p.~180]{Bo},
\cite[Thm~3.2, p.~34]{L} and Lemma~\ref{lemmaspec} below).
An important special case of a Krein space is when ${\cal{K}}$ is
positive except on a finite-dimensional subspace, i.e.
\begin{equation}
\kappa:= \dim {\cal{K}}_{-}\, < \infty.
\end{equation}
In this case the Krein space is called a \emph{Pontrjagin space} of index
$\kappa$.
Classical results of Pontrjagin (see~\cite[Thms~7.2 and 7.3, p.~200]{Bo}
and~\cite[p.~11-12]{L}) yield that any
selfadjoint operator $A$ on a Pontrjagin space is
definitizable, and that it has a $\kappa$-dimensional
negative subspace which is $A$-invariant.

We now explain how the abstract theory applies to the wave equation
in the Kerr geometry. In order to have a spectral theorem, the
Hamiltonian must be definitizable. There is no reason why $H$ should
be definitizable on the whole space $(r_1,\infty)\times S^2$, and this
leads us to consider the wave equation in ``finite volume''
$[r_L,r_R]\times S^2$ with Dirichlet boundary conditions. Thus setting
$\Psi =(\Phi,i \Phi_t)$ and regarding the two components $(\Psi_{1},\Psi_{2})$
of $\Psi$ as independent
functions, we consider the vector space
${\cal{P}}_{r_L,r_R}=(H^{1,2}\oplus L^2)([r_L,r_R]\times S^2)$ with
Dirichlet boundary conditions
\begin{equation}\label{Dirichlet}
\Psi_1(r_L) \;=\; 0 \;=\; \Psi_1(r_R)\:.
\end{equation}
Our definition of~$H^{1,2}([r_L,r_R]\times S^2)$ coincides with that
of the space~$W^{1,2}((r_L,r_R) \times S^2)$
in~\cite[Section~7.5]{GT}. Note that we only impose boundary
conditions on the first component~$\Psi_{1}$ of $\Psi$, which lies
in~$H^{1,2}$. According to the trace theorem~\cite[Part II, Section
5.5, Theorem~1]{Evans}, the boundary values of a function
in~$H^{1,2}([r_L, r_R] \times S^2)$ are in $L^2(S^2)$, and therefore
we can impose Dirichlet boundary conditions. We endow this vector
space with the inner product associated to the energy; i.e.\ in
analogy to~(\ref{energysp}),
\begin{eqnarray}
\lefteqn{ \bra \Psi, \Psi'\ket \;=\; \int_{r_L}^{r_R}
dr\int_{-1}^{1} d(\cos \vartheta) \left\{
\left({\frac{(r^{2}+a^{2})^{2}}{\Delta}} - a^{2}\,\sin^{2}\vartheta \right) \overline{\Psi_2}\, \Psi_2' \right. \nonumber } \\
&& \left. +\Delta \, \overline{\partial_{r}\Psi_1}
\,\partial_{r}{\Psi}_1' +\sin^{2}\vartheta
\,\overline{\partial_{\cos \vartheta}\Psi_1}\,\partial_{\cos
\vartheta}{\Psi}_1'  +\left(
{\frac{1}{\sin^{2}\vartheta}}-{\frac{a^{2}}{\Delta}}\right)\,\overline{\partial_{\varphi}\Psi_1}
\,\partial_{\varphi}{\Psi}_1' \right\} . \label{energyscalar}
\end{eqnarray}

\begin{Lemma}\label{lemma31h}
For every $r_R>r_1$ there is a countable set $E\subset (r_1,r_R)$ such
that for all $r_L \in (r_1,r_R)\setminus E$, the inner product space
${\cal{P}}_{r_L,r_R}$ is a Pontrjagin space. The topology
on~${\cal{P}}_{r_L,r_R}$ is the same as that on~$(H^{1,2} \oplus L^2)([r_L, r_R] \times S^2)$.
\end{Lemma}
{\Proof} Since~(\ref{energyscalar}) involves no terms which mix the
first component of $\Psi$ with the second component, ${\cal{P}}_{r_L,r_R}$ clearly has an
orthogonal direct sum decomposition ${\cal{P}}_{r_L,r_R} = V_1\oplus
V_2$ with $V_{1\!/\!2}=\{\Psi \in{\cal{P}}_{r_L,r_R} \;:\;
\Psi_{2\!/\!1}\equiv0\}$.
Furthermore, it is obvious that the space $(V_2,\bra ., .\ket \,)$
has a positive scalar product and
that the corresponding norm is equivalent to the
$L^2$-norm. Hence it remains to consider $V_1$, i.e. the space
$H^{1,2}([r_L,r_R]\times S^2)$ with Dirichlet boundary conditions and
the inner product
\begin{eqnarray}
\bra \Phi, \Phi'\ket &=& \int_{r_L}^{r_R}
dr \int_{-1}^{1} d(\cos \vartheta) \nonumber \\
&& \times \left\{ \Delta \, \overline{\partial_{r}\Phi} \,\partial_{r}\Phi' +\sin^{2}\vartheta
\,\overline{\partial_{\cos \vartheta}\Phi}\,\partial_{\cos \vartheta}\Phi'
+\left( {\frac{1}{\sin^{2}\vartheta}}-{\frac{a^{2}}{\Delta}}\right)\,
k^2\:\overline{\Phi} \Phi' \right\} . \label{3ip}
\end{eqnarray}
Transforming to the variable $u$, (\ref{51a}), and using the
representation~(\ref{ESP}), one sees that on the subspace
$C^2([u_L, u_R] \times S^2)$ the inner product~(\ref{3ip}) can be written as
\beq \label{sprep}
\bra \Phi\:,\: \Phi' \ket \;=\; (\Phi, A\Phi')_{L^2([u_L, u_R] \times S^2, d\mu)}
\eeq
with $A$ according to~(\ref{Adef}). Here we set $u_L=u(r_L)$, $u_R=u(r_R)$, and
$d\mu$ is the measure~(\ref{mudef}).
$A$ is a Schr{\"o}dinger operator with smooth potential on a compact
domain. Standard elliptic results~\cite[Proposition~2.1 and
the remark before Proposition~2.7]{TII} yield that
$H$ is essentially selfadjoint in the Hilbert space ${\cal{H}} =
L^2([u_L,u_R]\times S^2, d\mu)$. It has a purely discrete
spectrum which is bounded from below and has no limit points.
The corresponding eigenspaces are finite-dimensional, and the
eigenfunctions are smooth.

Let us analyze the kernel of $A$. Separating and using that the
Laplacian on $S^2$ has eigenvalues $-l(l+1)$, $l\in\mathbb{N}_0$, $A$
has a non-trivial kernel if and only if for some $l\in\mathbb{N}_0$, the solution
of the ODE
\beq \label{37x}
\left[-\frac{\partial}{\partial u} (r^2+a^2)
\frac{\partial}{\partial u} + \frac{\Delta}{r^2+a^2}\: l(l+1)
- \frac{a^2 k^2}{r^2+a^2} \right] \phi(u)=0
\eeq
with boundary conditions $\phi(u_R)=0$ and $\phi'(u_R)=1$ vanishes at
$u=u_L$. Since this $\phi$ has at most a countable number of zeros on
$(-\infty ,u_R]$ (note that $\phi(u)=0$ implies $\phi'(u)\not= 0$ because
otherwise $\phi$ would be trivial), $\phi$ vanishes at $u_L$ only if
$u_L\in E_l$ with $E_l$ countable. We conclude that there is a
countable set $E= \cup_lE_l$ such that the kernel of $A$ is trivial
unless $u_L\in E$.

Assume that $u_L \notin E$. Then $A$ has no kernel, and so we can
decompose ${\cal{H}}$ into the positive and negative spectral subspaces,
${\cal{H}} = {\cal{H}}_+\oplus {\cal{H}}_-$. Clearly, ${\cal{H}}_-$ is
finite-dimensional. Since its vectors are smooth functions, we can
consider ${\cal{H}}_-$ as a subspace of ${\cal{P}}_{r_L,r_R}$,
and according to~(\ref{sprep}) it is a negative subspace. Its
orthogonal complement in ${\cal{P}}_{r_L,r_R}$ is contained in
${\cal{H}}_+$ and is therefore positive. We conclude
that ${\cal{P}}_{r_L,r_R}$ is positive except on a finite-dimensional
subspace.

It remains to show that the topology induced by $\bra .\:,\: .\ket$ is
equivalent to the $H^{1,2}$-topology. Since on finite-dimensional
spaces all norms are equivalent, it suffices to consider for any
$\lambda_0>0$ the spectral subspace for $\lambda\geq\lambda_0$, denoted
by ${\cal{H}}_{\lambda_0}$. We choose $\lambda_0$ such that
$$1-\lambda_0 \;\leq\; V_0:=\min_{[r_L,r_R]}
\left( - \frac{a^2 k^2}{r^2+a^2} \right) <0\;.$$
Then for every $\Psi \in C^2 \cap {\cal{H}}_{\lambda_0}$,
\begin{eqnarray*}
\bra \Psi,\Psi \ket &=& \lbra \Psi,A \Psi \lket_{L^2(d\mu)}
 \;\stackrel{(\ast)}{\leq}\; c\: \|\Psi\|^2_{H^{1,2}} \\
\bra \Psi,\Psi \ket &\geq& \frac{1}{2}\: \lbra \Psi,A \Psi \lket_{L^2(d\mu)}
+\frac{\lambda_0}{2}\|\Psi\|^2_{L^2(d\mu)} \\
&\stackrel{(\ast)}{\geq}&
\frac{1}{2c}\: \|\Psi\|^2_{H^{1,2}} +\frac{V_0-1}{2} \: \|\Psi\|^2_{L^2(d\mu)}
+ \frac{\lambda_0}{2}\: \|\Psi\|^2_{L^2(d\mu)}
\;\geq\; \frac{1}{2c}\: \|\Psi\|^2_{H^{1,2}}\;,
\end{eqnarray*}
where in $(\ast)$ we used that the coefficients of the ODE~(\ref{37x}) are
bounded from above and below and that the zero order term is
bounded from below by $V_0$. \QED
We always choose $r_L$ and $r_R$ such that
${\cal{P}}_{r_L,r_R}$ is a Pontrjagin space and that our initial data is
supported in $[r_L,r_R] \times S^2$.

We now consider the Hamiltonian~(\ref{hamil}) on the Pontrjagin space
${\cal{P}}_{r_L, r_R}$ with domain $C^\infty([r_L, r_R] \times S^2)^2
\subset {\cal{P}}_{r_L, r_R}$. For clarity, we shall often denote this operator by~$H_{r_L, r_R}$.
\begin{Lemma} $H_{r_L, r_R}$ has a selfadjoint extension in~${\cal{P}}_{r_L, r_R}$.
\end{Lemma}
{\Proof} On the domain of $H$, the scalar product can be written
in analogy to~(\ref{ESP}) as
\[ \bra \Psi, \Psi' \ket \;=\; (\Psi, S \Psi')_{L^2([u_L, u_R] \times S^2,
d\mu)} \:, \]
where the operator $S$ acts on the two components of $\Psi$ as the
matrix
\begin{equation}
S=\left( \begin{array}{cc}
A & 0 \\
0 & 1
\end{array} \right) ,
\end{equation}
where~$A$ is again given by~(\ref{Adef}) and $d\mu$ is the measure~(\ref{mudef}).
As shown in Lemma~\ref{lemma31h}, $S$ has a selfadjoint extension and
is invertible. We introduce on~$C^\infty_0((u_L, u_R) \times S^2)^2$
the operator~$B$ by $B=|S|^{-\frac{1}{2}} S
H |S|^{-\frac{1}{2}}$. The fact that $H$ is symmetric in ${\mathcal{P}}_{r_L, r_R}$
implies that~$B$ is symmetric in~$L^2([u_L, u_R] \times S^2, d\mu)$. 
A short calculation shows that
\[ B^2 \;=\; \left( \begin{array}{cc}
|A| & |A|^{-\frac{1}{2}} A \beta \\
\beta |A|^{-\frac{1}{2}} A & |A| +\beta^2 \end{array} \right) . \]
Treating the terms involving~$\beta$ as a relatively compact perturbation,
we readily find that~$B^2$ is selfadjoint on~$L^2([u_L, u_R] \times S^2, d\mu)$
with domain~${\mathcal{D}}(B^2) = {\mathcal{D}}(A) \oplus {\mathcal{D}}(A)$.
Consequently, the spectral calculus gives us a selfadjoint extension
of~$B$ with domain~${\mathcal{D}}(B) = {\mathcal{D}}(A^{\frac{1}{2}})\oplus
{\mathcal{D}}(A^{\frac{1}{2}})$.
We extend~$H$ to the domain ${\mathcal{D}}(H) := |S|^{-\frac{1}{2}} \:{\mathcal{D}}(B)$.

We now show that with this new domain, $H$ is selfadjoint
on~${\mathcal{P}}_{r_L, r_R}$.
Suppose that for some vectors~$\Psi, \Phi \in {\mathcal{P}}_{r_L, r_R}$,
\[ \bra \Psi, H \Psi' \ket \;=\; \bra \Phi, \Psi' \ket \qquad
{\mbox{for all $\Psi' \in {\mathcal{D}}(H)$}}\:. \]
It then follows that
\[ (|S|^{\frac{1}{2}} \Psi, B\: |S|^{\frac{1}{2}} \Psi')_{L^2(d\mu)}
\;=\; (S\,|S|^{-\frac{1}{2}} \Phi, |S|^{\frac{1}{2}} \Psi')_{L^2(d\mu)} \qquad
{\mbox{for all $\Psi' \in {\mathcal{D}}(H)$}} \]
(note that the vectors $\Psi, \Phi \in {\mathcal{P}}_{r_L, r_R}$
lie in $H^{1,2} \oplus L^2$ and
thus both~$|S|^{\frac{1}{2}} \Psi,  S |S|^{-\frac{1}{2}} \Phi \in L^2(d\mu)$).
By definition of the domain of~$H$, this is equivalent to
\[ (|S|^{\frac{1}{2}} \Psi, B\: \tilde{\Psi})_{L^2(d\mu)}
\;=\; (S\,|S|^{-\frac{1}{2}} \Phi, \tilde{\Psi})_{L^2(d\mu)} \qquad
{\mbox{for all $\tilde{\Psi} \in {\mathcal{D}}(B)$}}\:. \]
Since~$B$ is selfadjoint, it follows that the vector~$|S|^{\frac{1}{2}} \Psi$
lies in the domain of~$B$ and that $B\:|S|^{\frac{1}{2}} \Psi = S\,|S|^{-\frac{1}{2}} \Phi$.
This implies that~$\Psi \in {\mathcal{D}}(H)$ and that~$H \Psi = \Phi$.
\QED

We now prove a basic lemma on the structure of the spectrum of the
Hamiltonian $H_{r_L,r_R}$.
\begin{Lemma} \label{lemmaspec}
The spectrum of $H_{r_L,r_R}$ is purely discrete. It consists
of finitely many complex spectral points appearing as complex
conjugate pairs, and of an infinite sequence of real eigenvalues
with no accumulation points.
\end{Lemma}
{\Proof}
Since the operator $H_{r_L,r_R}$ is essentially selfadjoint, there exists a
negative definite subspace $L_{-}$ of ${\cal P}_{r_L,r_R}$ of dimension $\kappa$ which
is $H_{r_L, r_R}$-invariant (see~\cite[p.\ 11]{L}).
Let $p_{0}$ denote the minimal polynomial of $H_{r_L, r_R}$ on~$L_{-}$, i.e.
\[ p_{0}(H_{r_L,r_R})\,L_{-} \;=\; 0 \]
with $\deg p_{0}\leq \kappa$ minimal. Furthermore, we let $p$ be the real polynomial of
degree~$\leq 2\kappa$ defined by~$p=p_{0}\,\overline{p_{0}}$.
We claim that ${\mbox {im}}\, p\,(H_{r_L,r_R})$ is a positive semi-definite
subspace. Indeed, we have for all $x \in {\cal P}_{r_L,r_R}$,
\begin{equation}
\bra \overline{p_{0}}\,(H_{r_L,r_R})\,x,L_{-}\ket \;=\; \bra x, p_{0}\,(H_{r_L,r_R})L_{-}\ket
\;=\; 0,
\end{equation}
so that
\begin{equation}
{\mbox {im}}\,p\,(H_{r_L,r_R}) \;\subset\; {\mbox {im}}\,{{\overline p}_{0}}\,(H_{r_L,r_R})
\;\subset\; ({L_{-}})^{\perp} \;\subset\; ({\cal P}_{r_L,r_R})_{+},
\end{equation}
as claimed.

Next, since the square of the operator $H_{r_L,r_R}$ is elliptic, it follows that
\begin{equation}
\dim {\mbox{ker}}\,(p\,(H_{r_L,r_R}) < \infty,
\end{equation}
and from~\cite[Prp.~2.1]{L}, we know that for each eigenvalue $\xi$ of $H_{r_L,r_R}$,
the corresponding Jordan chain has finite length bounded by $2 \kappa + 1$.
It follows that $p^{2\kappa +1}(H_{r_L,r_R})$ has a finite-dimensional kernel and
no Jordan chains. This implies that
\begin{equation}
{\mbox {im}}\,p^{2\kappa +1}\,(H_{r_L,r_R}) \cap  {\mbox{ker}}\,p^{2\kappa +1}\,(H_{r_L,r_R})
\;=\; \{0\} \:.
\end{equation}
Furthermore, since the operator $p^{2k+1}(H_{r_L, r_R})$ is selfadjoint, its
image and kernel are clearly orthogonal.

The image of $p^{2k+1}(H_{r_L, r_R})$ is contained in~${\mbox {im}}\,p\,(H_{r_L,r_R})$
and is therefore positive semi-definite. We shall now show that the space
${\mbox {im}}\,p^{2\kappa +1}\,(H_{r_L,r_R})$ is actually positive definite.
To this end, we let $N$ be its null space,
\begin{equation}
N:=\{x\in {\mbox {im}}\,p^{2\kappa +1}\,(H_{r_L,r_R}), \,\bra x\:, \:x\ket=0\} .
\end{equation}
For all $x\in N$ and $y \in {\cal{D}}(H_{r_L,r_R})$, we have
\begin{equation}
\bra x\:,\:p^{2\kappa +1}\,(H_{r_L,r_R})\,y \ket =0,
\end{equation}
which is equivalent to
\begin{equation}
\bra p^{2\kappa +1}\,(H_{r_L,r_R})\,x\:,\:y \ket = 0,
\end{equation}
because $p$ is real. Since the scalar product is non-degenerate,
this implies that
\begin{equation}
p^{2\kappa +1}\,(H_{r_L,r_R})\,x = 0\:.
\end{equation}
But we have just shown that
${\mbox{ker}}\,p^{2\kappa +1}\,(H_{r_L,r_R})$ and ${\mbox {im}}\,p^{2\kappa +1}\,(H_{r_L,r_R})$
have trivial intersection. It follows that~$x=0$ and therefore that
${\mbox {im}}\,p^{2\kappa +1}\,(H_{r_L,r_R})$ is positive definite, as claimed.

Restricting $H_{r_L, r_R}$ to ${\mbox {im}}\,p^{2\kappa +1}\,(H_{r_L,r_R})$, we have a
selfadjoint operator on a Hilbert space. Thus the spectral theorem in
Hilbert space applies, and the ellipticity of $H_{r_L, r_R}^2$ yields that the
spectrum is purely discrete.
On the finite-dimensional orthogonal complement
${\mbox{ker}}\,(p^{2\kappa +1}\,(H_{u_{a},u_{b}}))$ we bring $H_{r_L, r_R}$ into the
Jordan canonical form. \QED

\section{Resolvent Estimates} \label{sec5}
\setcounter{equation}{0}
In this section we consider the Hamiltonian $H$ as a non-selfadjoint operator
on the Hilbert space~${\cal{H}}$ with the scalar product~$(.,.)$ according
to~(\ref{psp}). We work either in infinite volume with domain of definition
${\cal{D}}(H) = C^\infty_0((r_1, \infty) \times S^2)^2$ or in the finite box
$r \in [r_L, r_R]$ with domain of definition given by the functions
in~$C^\infty((r_L, r_R) \times S^2)^2$ which satisfy the boundary conditions~(\ref{Dirichlet}).
Some estimates will hold in the same way in finite and infinite volume.
Whenever this is not the case, we distinguish between finite and
infinite volume with the subscripts $_{r_L, r_R}$ and $_\infty$, respectively.
We always consider a fixed $k$-mode.

The next lemma shows that the operator $H-\omega$ is invertible if either
$|{\mbox{Im}}\, \omega|$ is large or $|{\mbox{Im}}\, \omega| \neq 0$
and $|{\mbox{Re}}\, \omega|$ is large. The second case is more subtle,
and we prove it using a spectral decomposition of the elliptic operator $A$
which generates the energy scalar product.
This lemma will be very useful in Section~\ref{sec8}, because it will make
it possible to move the contour integrals so close to the real axis that
the angular estimates of Lemma~\ref{lemmaangular} apply.
By a slight abuse of notation we use the same notation for $H$ and its
closed extension.

\begin{Lemma} \label{lemma52imp}
There are constants $c, K>0$ such that for all $\Psi \in {\cal{D}}(H)$
and $\omega \in \C$,
\[ \| (H-\omega) \Psi \| \;\geq\; \frac{1}{c} \left(|{\mbox{\rm{Im}}}\,\omega|
- \frac{K}{1 + |{\mbox{\rm{Re}}}\, \omega|} \right) \| \Psi \| \:. \]
\end{Lemma}
{\Proof}
For every unit vector $\Psi \in {\cal{D}}(H)$,
\begin{eqnarray}
\|(H-\omega) \Psi\| &\geq& |(\Psi, (H-\omega) \Psi)| \;\geq\;
|{\mbox{Im}}\, (\Psi, (H-\omega) \Psi)| \nonumber \\
&\geq& |{\mbox{Im}}\, \omega| \:-\:
\frac{1}{2} \left|(\Psi, (H-H^*) \Psi) \right| . \label{Hoest}
\end{eqnarray}
It is useful to work again in the variable $u$ and the representation
(\ref{PSP}) of the scalar product $(.,.)$
on $C^2_0(\R \times S^2)^2$.
We introduce on $C^2_0(\R \times S^2)^2$ the operator $H_+$ by
\[ H_+ \;=\; \left( \begin{array}{cc} 0 & 1 \\ A+\delta & \beta \end{array} \right) . \]
Comparing with~(\ref{PSP}) one sees that $H_+$ is formally selfadjoint
w.r.\ to the scalar product $(.,.)$. Furthermore, one sees from~(\ref{HAb}) that
$H_+$ differs from $H$ only by a bounded operator,
\[ \|H - H_+\| \;=\; \left\| \left( \begin{array}{cc} 0 & 0 \\ -\delta & 0 \end{array}
\right) \right\| \;\leq\; c \:. \]
Thus on $C^2(\R \times S^2)^2$,
\begin{eqnarray*}
\| H - H^* \| &=& \|(H-H_+) - (H-H_+)^*\| \\
&\leq& \|H-H_+\| + \|(H-H_+)^*\|
\;=\; 2\:\|H-H_+\| \;\leq\; 2c \:,
\end{eqnarray*}
and substituting this bound into~(\ref{Hoest}), we conclude that
\[ \|(H-\omega) \Psi\| \;\geq\; \left(|{\mbox{Im}}\, \omega| - c \right)
\|\Psi\| \:. \]
In view of this inequality it remains to consider the case
where $|{\mbox{Re}}\, \omega|$ is large.

Using standard elliptic theory (see again \cite[p.~86, Proposition~2.7]{TII}
and~\cite{Ch})),
the operator $A$ with domain ${\cal{D}}(A) =
C^\infty_0(\R \times S^2)$ is essentially selfadjoint on the Hilbert space
$L^2(d\mu) := L^2(\R \times S^2, d\mu)$, with $d\mu$ according to~(\ref{mudef}).
Clearly, $A$ is bounded from below, $A \geq -c$,  and thus $\sigma(A)
\subset [-c,\infty)$.
For given $\Lambda \gg 1$ we let
$P_0$ and $P_\Lambda$ be the spectral projectors corresponding to
the sets $[-c, \Lambda^2)$ and $[\Lambda^2, \infty)$, respectively. We decompose
a vector $\Psi \in {\cal{H}}$ in the form $\Psi = \Psi_0 + \Psi_\Lambda$ with
\[ \Psi_0 \;=\; \left( \begin{array}{cc} P_0 & 0 \\ 0 & P_0 \end{array} \right) \Psi \;,\qquad
\Psi_\Lambda \;=\; \left( \begin{array}{cc} P_\Lambda & 0 \\ 0 & P_\Lambda
\end{array} \right) \Psi \:. \]
This decomposition is orthogonal w.r.\ to the energy scalar product,
\[ \bra \Psi_\Lambda, \Psi_0 \ket \;=\; \lbra \Psi_\Lambda,\: \left( \begin{array}{cc} A & 0 \\ 0 & 1
\end{array} \right) \Psi_0 \lket_{L^2(d\mu)} \;=\; 0 \:. \]
However, our decomposition is {\em{not}} orthogonal w.r.\ to the scalar product
$(.,.)$, because
\[ ( \Psi_\Lambda, \Psi_0 ) \;=\; \lbra \Psi_\Lambda,\: \left( \begin{array}{cc} A+\delta & 0 \\ 0 & 1
\end{array} \right) \Psi_0 \lket_{L^2(d\mu)}
\;=\; \lbra \Psi_\Lambda,\: \left( \begin{array}{cc} \delta & 0 \\ 0 & 0
\end{array} \right) \Psi_0 \lket_{L^2(d\mu)} \:. \]
But at least we obtain the following inequality,
\beq \label{rel2}
|( \Psi_\Lambda, \Psi_0 ) | \;\leq\;
c\: \|\Psi_0\| \: \|\Psi_\Lambda^1 \|_{L^2(d\mu)}\:,
\eeq
where $\Psi^1_\Lambda$ denotes the first component of $\Psi_\Lambda$.
Using that
\[ \|\Psi_\Lambda^1 \|^2_{L^2(d\mu)} \;=\;
\bra \Psi_\Lambda^1, A^{-1} \Psi_\Lambda^1 \ket
\;\leq\; \frac{1}{\Lambda^2}\: \|\Psi_\Lambda\|^2 \;, \]
we can also write~(\ref{rel2}) in the more convenient form
\[ |( \Psi_\Lambda, \Psi_0 ) | \;\leq\;
\frac{c}{\Lambda}\; \|\Psi_0\| \: \|\Psi_\Lambda\| \:. \]
Choosing $\Lambda$ sufficiently large, we obtain
\[ \|\Psi\|^2 \;=\; \|\Psi_\Lambda\|^2 \:+\: 2\: {\mbox{Re}}\,
( \Psi_\Lambda, \Psi_0) \:+\: \|\Psi_0\|^2 \;\leq\; 4 \left(
\|\Psi_\Lambda\| + \|\Psi_0\| \right)^2 \]
and thus
\beq \label{rel3}
\|\Psi\| \;\leq\; 2 \left( \|\Psi_\Lambda\| + \|\Psi_0\| \right) .
\eeq
Furthermore, we can arrange by choosing $\Lambda$ sufficiently large
that
\[ \bra \Psi_\Lambda, \Psi_\Lambda \ket \;=\;
\lbra \Psi_\Lambda, \left( \begin{array}{cc} A & 0 \\ 0 & 1
\end{array} \right) \Psi_\Lambda \lket_{L^2(d\mu)}
\;\geq\; \frac{1}{2} \: \lbra \Psi_\Lambda, \left( \begin{array}{cc} A+\delta
& 0 \\ 0 & 1 \end{array} \right) \Psi_\Lambda \lket_{L^2(d\mu)}
\;=\; \frac{1}{2}\: \|\Psi_\Lambda\|^2. \]

Next we estimate the inner products $\bra \Psi_\Lambda, H \Psi_0 \ket$,
$(\Psi_0, H \Psi_\Lambda)$ and $(\Psi_0, H \Psi_0)$. The calculations
\begin{eqnarray*}
\bra \Psi_\Lambda, H \Psi_0 \ket &=& \lbra \Psi_\Lambda,
\left( \begin{array}{cc} 0 & A \\ A & \beta
\end{array} \right) \Psi_0 \lket_{L^2(d\mu)}
\;=\; \lbra \Psi_\Lambda,
\left( \begin{array}{cc} 0 & 0 \\ 0 & \beta
\end{array} \right) \Psi_0 \lket_{L^2(d\mu)} \\
( \Psi_0, H \Psi_\Lambda ) &=& \lbra \Psi_0,
\left( \begin{array}{cc} 0 & A+\delta \\ A & \beta
\end{array} \right) \Psi_\Lambda \lket_{L^2(d\mu)}
\;=\; \lbra \Psi_0,
\left( \begin{array}{cc} 0 & \delta \\ 0 & \beta
\end{array} \right) \Psi_\Lambda \lket_{L^2(d\mu)} \\
|(\Psi_0, H \Psi_0)| &=& \left| \lbra \Psi_0,
\left( \begin{array}{cc} 0 & A+\delta \\ A & \beta
\end{array} \right) \Psi_0 \lket_{L^2(d\mu)} \right| \\
&\leq& c\: \|\Psi_0\|_{L^2(d\mu)} + 2\: \|A \Psi_0^1\|_{L^2(d\mu)}\:
\|\Psi_0^2\|_{L^2(d\mu)} \\
\|A \Psi_0^1\|_{L^2(d\mu)}^2 &=& \lbra \Psi_0, \left( \begin{array}{cc}
A^2 & 0 \\ 0 & 0 \end{array} \right) \Psi_0 \lket_{L^2(d\mu)} \\
&\leq& \Lambda^2\:
\lbra \Psi_0, \left( \begin{array}{cc}
A & 0 \\ 0 & 1 \end{array} \right) \Psi_0 \lket_{L^2(d\mu)}
\;=\; \Lambda^2\: \|\Psi_0\|^2
\end{eqnarray*}
give us the bounds
\begin{eqnarray*}
|<\!\! \Psi_\Lambda, H \Psi_0 \!\!>| &\leq& c\: \|\Psi_\Lambda\|\: \|\Psi_0\| \\
|( \Psi_0, H \Psi_\Lambda )| &\leq& c\: \|\Psi_0\|\: \|\Psi_\Lambda\| \\
|( \Psi_0, H \Psi_0 )| &\leq& (c+2 \Lambda)\: \|\Psi_0\|^2\:.
\end{eqnarray*}

Using the above inequalities, we can estimate the inner product
$\bra \Psi_\Lambda, (H-\omega) \Psi \ket$ by
\begin{eqnarray*}
|\bra \Psi_\Lambda, (H-\omega) \Psi \ket| &\geq&
|\bra \Psi_\Lambda, (H-\omega) \Psi_\Lambda \ket| -
|\bra \Psi_\Lambda, (H-\omega) \Psi_0 \ket| \\
&\geq& \frac{|{\mbox{Im}}\, \omega|}{2}\: \|\Psi_\Lambda\|^2
- c\: \|\Psi_0\|\: \|\Psi_\Lambda\|\:.
\end{eqnarray*}
Applying the Cauchy-Schwarz inequality $|\bra \Psi_\Lambda, (H-\omega) \Psi \ket| \leq
c_1 \:\|\Psi_\Lambda\|\: \|(H-\omega) \Psi\|$ and dividing by
$\|\Psi_\Lambda\|$, we obtain (possibly after increasing $c$) that
\beq \label{es1}
\|(H-\omega) \Psi \| \;\geq\; \frac{|{\mbox{Im}}\, \omega|}{c}\: \|\Psi_\Lambda\|
\:-\: \|\Psi_0\|\:.
\eeq
Next we estimate the inner product $(\Psi_0, (H-\omega) \Psi)$,
\begin{eqnarray*}
|( \Psi_0, (H-\omega) \Psi )| &\geq&
|(\Psi_0, (H-\omega) \Psi_0 )| -
|(\Psi_0, (H-\omega) \Psi_\Lambda )| \\
&\geq& (|\omega| - c- 2 \Lambda)\: \|\Psi_0\|^2 - c
\left(1+\frac{|\omega|}{\Lambda} \right) \|\Psi_0\|\: \|\Psi_\Lambda\| \:.
\end{eqnarray*}
We apply the Cauchy-Schwarz inequality $( \Psi_0, (H-\omega) \Psi ) \leq
\|\Psi_0\|\: \|(H-\omega) \Psi\|$ and divide by $\|\Psi_0\|$,
\beq \label{es2}
\|(H-\omega) \Psi\| \;\geq\; (|\omega| - c- 2\Lambda)\: \|\Psi_0\|
- c \left(1 + \frac{|\omega|}{\Lambda} \right) \|\Psi_\Lambda\| \:.
\eeq
Choosing $\Lambda = (|\omega|-c)/4$ and increasing $c$,
the inequalities~(\ref{es1}) and~(\ref{es2})
give for sufficiently large $|\omega|$ the bounds
\begin{eqnarray*}
\|(H-\omega) \Psi\| &\geq& \frac{|{\mbox{Im}}\, \omega|}{c}\: \|\Psi_\Lambda\|
\:-\: \|\Psi_0\| \\
\|(H-\omega) \Psi\| &\geq& \frac{|\omega|}{2}\: \|\Psi_0\|
- c\: \|\Psi_\Lambda\| \:.
\end{eqnarray*}
Multiplying the second inequality by $4/|\omega|$ and adding the
first inequality, we conclude that
\[ 2\: \|(H-\omega) \Psi\| \;\geq\; \left(\frac{|{\mbox{Im}}\, \omega|}{c}
- \frac{4c}{|\omega|} \right) \|\Psi_\Lambda\| + \|\Psi_0\| \:. \]
The result now follows from~(\ref{rel3}).
\QED

With the last lemma at hand, we are ready to introduce the resolvent. Namely, we let
\beq \label{Odef}
\Omega \;=\; \left\{ \omega \in \C \::\:
|{\mbox{Im}}\, \omega| \;\geq\; \frac{2K}{1 + |{\mbox{Re}}\, \omega|} \right\}
\eeq
with $K$ as in Lemma~\ref{lemma52imp}.
\begin{Corollary} \label{cor53}
If $\omega \in \Omega$, the operator $H-\omega$ is invertible. The corresponding
{\bf{resolvent}}
\[ S(\omega) \;:=\; (H-\omega)^{-1} \]
satisfies the bound
\begin{equation} \label{reses}
\|S(\omega)\| \;\leq\; \frac{c}{|{\mbox{\rm{Im}}}\, \omega|}
\end{equation}
with $c$ independent of $\omega \in \Omega$.
\end{Corollary}
{\Proof}
In view of the preceding lemma, it suffices to show that the image
of $H-\omega$ is dense in $\cal H$ for any $\omega \in \Omega$. Otherwise, there would exist
a non-zero $\hat{\Psi} \in \cal{H}$ such that
\begin{equation}
\bra (H-\omega)\Psi,\hat{\Psi}\ket \;=\; 0 \qquad {\mbox{for all $\Psi \in {\cal D}(H)$}} \:,
\end{equation}
that is a weak solution of the equation
$(H-\omega)\Psi=0$. By the regularity theorem for elliptic operators on
manifolds with boundary (cf. \cite[Chapter5, Theorem~1.3]{TI}), every weak solution
of this equation is a solution in the strong sense. These have been
ruled out by the preceding Lemma.
\QED
Using~(\ref{Odef}) in~(\ref{reses}), we immediately get the bound
\beq
\|S(\omega)\| \;\leq\; c\: (1+|{\mbox{Re}}\, \omega|)\:. \label{Sbd2}
\eeq
Since $S(\omega)$ is a bounded operator, its domain of definition can clearly
be chosen to be the whole Hilbert space.
We shall assume until the end of this section that $\omega \in \Omega$.

The next lemma gives detailed estimates for the
difference of the resolvents $S_{r_L, r_R}$ and
$S_\infty$ in finite and infinite volume, respectively.
By $Q_\lambda(\omega)$ we denote a given projector onto an
invariant subspace of the angular operator ${\cal{A}}_\omega$
corresponding to the spectral parameter $\lambda$  of dimension
at most $N$ (see Lemma~\ref{lemmaangular} for details).
\begin{Lemma} \label{lemma54}
For every $\Psi \in C^\infty_0((r_L, r_R) \times S^2)^2$ and every $p \in \N$,
there is a constant $C=C(\Psi, p)$ (independent of $\omega$) such that
\begin{equation} \label{res1}
\left| \bra \Psi, \left[ S_{r_L, r_R}(\omega) - S_\infty(\omega)
\right] \Psi \ket \right|
\;\leq\; \frac{C}{1+|\omega|^p}\; \frac{1}{|{\mbox{\rm{Im}}}\, \omega|} \:.
\end{equation}
Furthermore, for every $\Psi \in C^\infty_0((r_L, r_R) \times S^2)^2$ and every
$p \in \N$ and $q \geq N$, there is a constant $C=C(\Psi, p, q)$
(independent of $\omega$ and $\lambda$) such that
\begin{equation} \label{res2}
\left| \bra \Psi, Q_\lambda \left[ S_{r_L, r_R}(\omega) - S_\infty(\omega)
\right] \Psi \ket \right|
\;\leq\; \frac{C}{(1+|\omega|^p)(1+|\lambda|^q)}\;
\frac{1}{|{\mbox{\rm{Im}}}\, \omega|}\; \|Q_\lambda \| \;.
\end{equation}
\end{Lemma}
{\Proof} By definition of the resolvent, $(H-\omega)\: S(\omega) \Psi=\Psi$. This
relation holds both in finite and in infinite volume, and thus
\[ \left((H-\omega) \left[S_{r_L, r_R}(\omega)-S_\infty(\omega) \right] \Psi \right)
(r, \vartheta) \;=\; 0
\spc {\mbox{if $r_L \leq r \leq r_R$}}. \]
Iterating this identity and using the fact that $H$ and $S$ commute, we see
that on $[r_L, r_R] \times S^2$,
\begin{equation} \label{hes}
\omega^{p+1} \left[S_{r_L, r_R}(\omega)-S_\infty(\omega) \right] \Psi \;=\;
\left[S_{r_L, r_R}(\omega)-S_\infty(\omega) \right] H^{p+1} \:\Psi \:.
\end{equation}
Combining this identity with the Schwarz-type inequality~(\ref{schwarz}), we
obtain
\begin{eqnarray*}
\left| \bra \Psi, \left[S_{r_L, r_R}(\omega)-S_\infty(\omega) \right] \Psi \ket \right|
&\leq& c_1 \:\|S_{r_L, r_R}(\omega)-S_\infty(\omega)\|\: \|\Psi\|^2 \\
|\omega^{p+1}| \left| \bra \Psi, \left[S_{r_L, r_R}(\omega)-S_\infty(\omega) \right] \Psi \ket \right|
&\leq& \left| \bra \Psi, \left[S_{r_L, r_R}(\omega)-S_\infty(\omega) \right]
H^{p+1} \Psi \ket \right| \\
&\leq& c_1 \:\|S_{r_L, r_R}(\omega)-S_\infty(\omega)\|\: \|\Psi\|\:
\|H^{p+1} \Psi\| \:.
\end{eqnarray*}
Since $\Psi$ is smooth and has compact support, $H^{p+1} \Psi$ also has these properties.
The estimate~(\ref{Sbd2}) gives~(\ref{res1}).

In order to prove~(\ref{res2}), we first combine~(\ref{hes})
with~(\ref{schwarz}) to obtain
\begin{eqnarray}
\lefteqn{ (1+|\omega|^{p+1}) \left| \bra \Psi, Q_\lambda
(S_{r_L, r_R}-S_\infty) \Psi \ket \right| } \nonumber \\
&\leq& c_1 \:\|Q_\lambda \|\:
\|S_{r_L, r_R}-S_\infty\| \:\|\Psi\| \left(
\| \Psi \| + \| H^{p+1} \Psi\| \right). \label{nes1}
\end{eqnarray}
Since $q$ is at least as large as the dimension of the invariant subspace
corresponding to $\lambda$, $({\cal{A}}_\omega - \lambda)^q
Q_\lambda = 0$. Therefore, for every $\Psi' \in C^\infty_0((r_L, r_R) \times S^2)^2$,
\[ 0 \;=\; \bra \Psi, ({\cal{A}}_\omega - \lambda)^q Q_\lambda \Psi' \ket
\;=\; \bra ({\cal{A}}^*_\omega - \overline{\lambda})^q
\Psi, Q_\lambda \Psi' \ket \;. \]
Expanding the power $({\cal{A}}^*_\omega - \overline{\lambda})^q$
and using~(\ref{schwarz}), we obtain
\[ |\lambda|^q \left| \bra \Psi, Q_\lambda \Psi' \ket \right|
\;\leq\; \sum_{l=1}^{q} c_l\: |\lambda|^{q-l}
\left\| ({\cal{A}}_\omega^*)^l \Psi \right\| \left\|
\chi_{[r_L, r_R]}\: Q_\lambda \Psi' \right\| \]
with combinatorial factors $c_l$ (here~$\chi_{[r_L, r_R]}$ is the operator
of multiplication by the characteristic function).
Since the angular operator $\cal{A}_\omega^*$ is according to~(\ref{angular})
a polynomial in $\omega$ of degree two, the function
$({\cal{A}}_\omega^*)^l \Psi$ is also polynomial in $\omega$, i.e.
\[ ({\cal{A}}_\omega^*)^l \Psi \;=\; \sum_{p=0}^{2l} \omega^p\: \Psi_p \:, \]
where the functions $\Psi_p$ are composed of $\Psi$ and its angular
derivatives, as well as the coefficient functions of ${\cal{A}}_\omega^*$.
This gives the estimate
\[ \|({\cal{A}}_\omega^*)^l \Psi\| \;\leq\;
\sum_{p=0}^{2l} |\omega|^p\: \|\Psi_p\| \;\leq\;
c\:(1+|\omega|^{2l}) \]
with a constant $c$ which depends only on $\Psi$ and $l$.
We thus obtain
\[ |\lambda|^q \left| \bra \Psi, Q_\lambda \Psi' \ket \right|
\;\leq\; \sum_{l=1}^{q} c_l(\Psi)\: |\lambda|^{q-l}\: (1+|\omega|^{2l})
\left\|\chi_{[r_L, r_R]}\:Q_\lambda \Psi' \right\| . \]
Young's inequality allows us to compensate the lower powers of $\lambda$,
\[ |\lambda|^q \left| \bra \Psi, Q_\lambda \Psi' \ket \right|
\;\leq\; c(q, \Psi) \: (1+|\omega|^{2q}) \left\|\chi_{[r_L, r_R]}\:
Q_\lambda \Psi' \right\| . \]
We now choose $\Psi'$ equal to the left side of~(\ref{hes})
with $p=0$ and $p=r$ and take the sum of the resulting inequalities.
Applying again the Schwarz inequality, we obtain
\[ |\lambda|^q \:(1+|\omega|^r) \left| \bra \Psi, Q_\lambda
(S_{r_L, r_R}-S_\infty) \Psi \ket \right|
\;\leq\; c \: (1+|\omega|^{2q}) \:\|Q_\lambda\|\:
\|S_{r_L, r_R}-S_\infty\| \left(
\| \Psi \| + \| H^r \Psi\| \right). \]
By choosing $r$ sufficiently large, we can compensate the factor
$(1+|\omega|^{2q})$ on the right. More precisely,
\[ |\lambda|^q \:(1+|\omega|^{p+1}) \left| \bra \Psi, Q_\lambda
(S_{r_L, r_R}-S_\infty) \Psi \ket \right|
\;\leq\; c' \:\|Q_\lambda\|\:
\|S_{r_L, r_R}-S_\infty\| \left(
\| \Psi \| + \| H^{p+2q+1} \Psi\| \right). \]
Adding this inequality to~(\ref{nes1}) and
substituting the estimate~(\ref{Sbd2}) gives~(\ref{res2}).
\QED

\section{Separation of the Resolvent} \label{sec6}
\setcounter{equation}{0}
In this section we fix $\omega \not \in \sigma(H)$, so that the
resolvent $S=(H-\omega)^{-1}$ exists. As in the previous section,
we assume that $Q_\lambda$ is a given projector onto a finite-dimensional
invariant subspace of the angular operator ${\cal{A}}_\omega$ corresponding
to the spectral parameter $\lambda$.
Our goal is to represent the operator product $Q_\lambda S$ in terms
of the solutions of the radial ODE.

According to~(\ref{odes}) and~(\ref{radial}), the radial ODE is
\begin{equation}
\left[ -\frac{\partial}{\partial r} \Delta \frac{\partial}{\partial r} - \frac{(r^2+a^2)^2}{\Delta}\:
\left(\omega + \frac{a k}{r^2+a^2} \right)^2 + \lambda
\right] R(r) \;=\; 0 \:, \label{5a}
\end{equation}
where $\lambda$ is the separation constant.
We can assume that $k \geq 0$ because otherwise we reverse the sign of $\omega$.
We again work in the ``tortoise variable'' $u$, (\ref{51a}), and set
\beq \label{5ptrans1}
\phi(r) \;=\; \sqrt{r^2+a^2}\: R(r) \;.
\eeq
Then equation~(\ref{5a}) can be written as
\begin{equation}
\left[ \frac{1}{r^2+a^2}\: \frac{\partial}{\partial u}\: (r^2+a^2)\: \frac{\partial}{\partial u}
+ \left(\omega + \frac{a k}{r^2+a^2} \right)^2 - \frac{\lambda\: \Delta}{(r^2+a^2)^2} \right]
\frac{\phi}{\sqrt{r^2+a^2}} \;=\; 0 \;. \label{5b}
\end{equation}
Using that
\[ (r^2+a^2)\: \frac{\partial}{\partial u}\: (r^2+a^2)^{-\frac{1}{2}}
\;=\; -\frac{1}{2}\: (r^2+a^2)^{-\frac{1}{2}}\: \frac{\partial}{\partial u}(r^2+a^2)
\;=\; -\frac{\partial}{\partial u}\: (r^2+a^2)^{\frac{1}{2}} \;, \]
(\ref{5b}) simplifies to the Schr{{\"o}}dinger-type equation
\begin{equation}
\left(-\frac{\partial^2}{\partial u^2} + V(u) \right) \phi(u) \;=\; 0
\label{5ode}
\end{equation}
with the potential
\begin{equation} \label{5V}
V(u) \;=\; -\left( \omega + \frac{ak}{r^2+a^2} \right)^2 \:+\:
\frac{\lambda\:\Delta}{(r^2+a^2)^2} \:+\: \frac{1}{\sqrt{r^2+a^2}}\; \partial_u^2 \sqrt{r^2+a^2} \;.
\end{equation}

We let $\phi_1$ and $\phi_2$ be two solutions of~(\ref{5ode}) which are compatible
with the boundary conditions. More precisely, in finite volume we satisfy
the Dirichlet boundary conditions $\phi_1(u_L)=0$ and $\phi_2(u_R)=0$
(again with $u_L=u(r_L)$ and $u_R=u(r_R)$). Likewise, in infinite volume
we only consider the case ${\mbox{Im}}\, \omega<0$ and let $\phi_1$ and
$\phi_2$ be the fundamental solutions which decay exponentially
at $u=-\infty$ and $u=+\infty$,
respectively (the existence of these fundamental solution will be
established in Corollary~\ref{cor74}).
If the solutions $\phi_1$ and $\phi_2$ were linearly dependent, they would
give rise to a vector in the kernel of $H-\omega$, in contradiction to
our assumption $\omega \not \in \sigma(H)$.
Thus the Wronskian
\beq \label{wronski}
w(\phi_1, \phi_2) \;:=\; \phi_1'(u) \: \phi_2(u) - \phi_1(u) \: \phi_2'(u)
\eeq
is non-zero (note that $w$ is by definition independent of $u$).

We begin by constructing the ``Green's function'' corresponding to~(\ref{5ode}).
\begin{Lemma} \label{lemma61}
The function
\begin{equation} \label{sdef}
s(u,u') \;:=\; \frac{1}{w(\phi_1, \phi_2)} \:\times \left\{
\begin{array}{cl} \phi_1(u)\: \phi_2(u') & {\mbox{if $u \leq u'$}} \\
\phi_2(u)\: \phi_1(u') & {\mbox{if $u > u'$}} \end{array} \right.
\end{equation}
satisfies the distributional equation
\[ \left( -\frac{\partial^2}{\partial u^2} + V(u) \right) s(u,u') \;=\;
\delta(u-u')\;. \]
\end{Lemma}
{\Proof}
By definition of the distributional derivative,
\[ \int_{-\infty}^\infty \eta(u) \: (-\partial_u^2+V) s(u,u') \: du \;=\;
\int_{-\infty}^\infty \left((-\partial_u^2+V) \eta(u) \right) s(u,u') \: du \]
for every test function $\eta \in C^\infty_0(\R)$.
It is obvious from its definition that the function $s(.,u')$ is smooth except at the point $u=u'$, where its first derivative has a discontinuity. Thus after
splitting up the integral, we can integrate by parts twice to obtain
\begin{eqnarray*}
\lefteqn{ \int_{-\infty}^\infty \left((-\partial_u^2+V) \eta(u) \right) s(u,u') \: du } \\
&=& \int_{-\infty}^{u'} \eta(u) \: (-\partial_u^2+V) s(u,u') \: du
\:+\: \lim_{u \nearrow u'} \eta(u)\: \partial_u s(u,u') \\
&& +\int_{u'}^\infty \eta(u) \: (-\partial_u^2+V) s(u,u') \: du
\:-\: \lim_{u \searrow u'} \eta(u)\: \partial_u s(u,u') \:.
\end{eqnarray*}
Since for $u \neq u'$, $s$ is a solution of~(\ref{5ode}), the obtained integrals
vanish. Computing the limits with~(\ref{sdef}), we get
\begin{eqnarray*}
\lefteqn{ \int_{-\infty}^\infty \left((-\partial_u^2+V) \eta(u) \right) s(u,u') \: du
\;=\; \left( \lim_{u \nearrow u'} - \lim_{u \searrow u'} \right)
\eta(u)\: \partial_u s(u,u') } \\
&=& \frac{1}{w(\phi_1, \phi_2)}\: \eta(u') \left(
\phi_1'(u')\: \phi_2(u') - \phi_2'(u')\: \phi_1(u') \right) \;=\;
\eta(u')\:,
\end{eqnarray*}
where in the last step we used the definition of the Wronskian~(\ref{wronski}).
\QED
In what follows we also regard $s(u,u')$ as the integral kernel of
a corresponding operator $s$, i.e.
\[ (s \phi)(u) \;:=\; \int du'\: s(u,u')\: \phi(u')\: du'\:. \]

If $Q_\lambda$ projects onto an eigenspace of ${\cal{A}}_\omega$, we see
from~(\ref{odes}), (\ref{wavesep}), and~(\ref{5ptrans1}) that
\beq \label{sepgrn}
\square \left( (r^2+a^2)^{-\frac{1}{2}}\: Q_\lambda(\vartheta, \vartheta')\:
\: s(u, u') \right)
\;=\; (r^2+a^2)^{-\frac{1}{2}}\: Q_\lambda(\vartheta, \vartheta') \: \delta(u-u')\;.
\eeq
Loosely speaking, this relation means that the operator product
$Q_\lambda \,s$ is an angular mode of the Green's function of
the wave equation.
Unfortunately, $Q_\lambda$ might project onto an
invariant subspace of ${\cal{A}}_\omega$ which is {\em{not}} an eigenspace.
In this case, the angular operator has on the invariant subspace the
``Jordan decomposition''
\beq \label{Jordan}
{\cal{A}}_\omega \:Q_\lambda \;=\; (\lambda + {\cal{N}})\: Q_\lambda
\eeq
with ${\cal{N}}={\cal{N}}(\omega,\lambda)$ a nilpotent operator.
Lemma~\ref{lemma6g} extends~(\ref{sepgrn}) to this more general case.
In preparation, we need to consider powers of the operator $s$.
\begin{Lemma} \label{lemmaspower}
For every $l \in \N_0$, the operator $s^l$ is well-defined. Its
kernel $(s^l)(u,u')$ has regularity~$C^{2l-2}$.
\end{Lemma}
{\Proof} Writing out the operator products with the integral kernel,
one sees that the operator $s^l$ is obtained from $s$ by
iterated convolutions,
\beq \label{intit}
s^{p+1}(u,u') \;=\; \int s(u,u'')\: s^p(u'',u')\:du''\:.
\eeq
In the finite box, these convolution integrals are all finite because
$s(u,u')$ is continuous and the integration range is compact.
In infinite volume, the function $s(u,u')$ decays exponentially
as $u,u' \to \pm \infty$ (see Corollary~\ref{cor74}),
and so the integrals in~(\ref{intit}) are again finite.
Hence $s^l$ is well-defined.

Let us analyze the regularity of the integral kernel of $s^l$.
By definition, $s(u,u')$ is continuous, and~(\ref{intit}) immediately
shows that the same is true for $s^p(u,u')$. Differentiating
through~(\ref{intit}) and applying Lemma~\ref{lemma61}, one
sees that $s^p$ satisfies for $p>1$ the distributional equation
\[ \left( -\frac{\partial^2}{\partial u^2} + V(u) \right) (s^p)(u,u') \;=\;
(s^{p-1})(u-u')\;. \]
This shows that incrementing $p$ indeed increases the order of
differentiability by two.
\QED

\begin{Lemma} \label{lemma6g}
For given $\lambda \in \sigma({\cal{A}}_\omega)$ we let $g$ be the operator
\beq \label{gdef}
g \;=\; \sum_{l=0}^\infty (-{\cal{N}})^l\: s^{l+1} ,
\eeq
where ${\cal{N}}$ is the nilpotent matrix in the
Jordan decomposition~(\ref{Jordan}). Then
\beq \label{grn}
\square \left( (r^2+a^2)^{-\frac{1}{2}}\: Q_\lambda(\vartheta, \vartheta')\:
\: g(u, u') \right)
\;=\; (r^2+a^2)^{-\frac{1}{2}}\: Q_\lambda(\vartheta, \vartheta') \: \delta(u-u')\;.
\eeq
\end{Lemma}
Note that if $Q_\lambda$ projects onto an eigenspace, ${\cal{N}}$ vanishes and
thus $g=s$. Furthermore, since ${\cal{N}}$ is nilpotent, the
series in~(\ref{gdef}) is actually a finite sum. Thus in view of
Lemma~\ref{lemmaspower}, (\ref{gdef}) is indeed well-defined. \\[.5em]
{\em{Proof of Lemma~\ref{lemma6g}. }}
Denoting the radial operator with integral kernel $\delta(u-u')$ by $\1_u$, we can write the result of Lemma~\ref{lemma61} in the compact form
$(-\partial_u^2 + V)s=\1_u$. Hence on the invariant subspace, we
can do a Neumann series calculation,
\[ (-\partial_u^2 + V)\:g \;=\; \sum_{l=0}^\infty (-{\cal{N}})^l\:
(-\partial_u^2 + V) s^{l+1} \;=\; \sum_{k=0}^\infty (-{\cal{N}})^l\:
s^l \;=\; \1_u - {\cal{N}}\: g \;,\]
to obtain that $(-\partial_u^2 + V+ {\cal{N}})\:g = \1_u$.
According to~(\ref{odes}), (\ref{wavesep}), and~(\ref{5ptrans1}), this is equivalent to~(\ref{grn}).
\QED

We come to the separation of the resolvent.
In order to explain the difficulty, we point out that $H$ and $Q_\lambda$ do
not in general commute, and thus
\[ (H-\omega)\:Q_\lambda \;\neq\; Q_\lambda\: (H-\omega) \quad {\mbox{and}} \quad
Q_\lambda\:S \;\neq\; S\:Q_\lambda\;. \]
Therefore, one must be very careful with the orders of multiplication; in
particular, it is not possible to simplify the operator product
$(H-\omega) Q_\lambda S$. However, we know from the separation of variables
that for every solution $\Psi$ of the equation $(H-\omega) \Psi=0$, its projection $Q_\lambda(\omega) \Psi$ is again a solution.
In other words, $H$ and $Q_\lambda$ do commute on the kernel of $(H-\omega)$. This fact will be exploited in the proof of the
following proposition.

\begin{Prp} \label{thm63}
For $\omega \not \in \sigma(H)$ we let $Q_\lambda$ be a
spectral projector of the angular operator ${\cal{A}}_\omega$.
Then the resolvent of $H$ has the representation
\[ Q_\lambda\: S(\omega) \;=\; Q_\lambda\: T(\omega, \lambda) \;, \]
where $T$ is the operator with integral kernel
\begin{eqnarray}
\lefteqn{ T(u, \vartheta;u', \vartheta')
\;=\; \delta(\cos \vartheta-\cos \vartheta')\: \delta(u-u')\:
\left( \!\!\begin{array}{cc} 0 & 0 \\ 1 & 0 \end{array} \!\!\right)
} \nonumber \\
&&+ \delta(\cos \vartheta-\cos \vartheta')\: (r^2+a^2)^{-\frac{1}{2}}\:
g(u,u') \left( \!\!\begin{array}{cc}
\tau(u', \vartheta') & \sigma(u', \vartheta') \\
\omega\: \tau(u', \vartheta') &
\omega\: \sigma(u', \vartheta') \end{array} \!\!\right) .  \label{Tdef}
\end{eqnarray}
Here $g$ is the Green's function~(\ref{gdef}) (which depends
on $\omega$ and $\lambda$), and $\tau, \sigma$
are the functions
\begin{eqnarray*}
\sigma &=& (r^2+a^2)^{-\frac{3}{2}}
\left((r^2+a^2)^2 - \Delta\: a^2 \sin^2 \vartheta \right) \\
\tau &=& 2 a k \: (r^2+a^2)^{-\frac{3}{2}} \left( (r^2+a^2) - \Delta
\right) + \omega \sigma\: .
\end{eqnarray*}
\end{Prp}
{\Proof} Let us compute the operator product $(H-\omega) Q_\lambda T$.
We first consider the first summand in~(\ref{Tdef}), which we denote by $T_1$.
In this case, the operator product is
particularly simple because the second column in the matrix~(\ref{hamil})
involves no $u$- or $\vartheta$-derivatives. We obtain
\beq \label{T1eq}
((H-\omega)\: Q_\lambda \: T_1)(u,\vartheta; u',\vartheta') \;=\;
Q_\lambda(\vartheta, \vartheta')\: \delta(u,u') \left( \!\!\begin{array}{cc} 1 & 0 \\
\beta(u, \vartheta)-\omega & 0 \end{array} \!\!\right) .
\eeq

Next we consider the second summand in~(\ref{Tdef}), which we denote by $T_2$.
Fixing $u', \vartheta'$ and considering $T_2$ as a function of $u, \vartheta$,
we see from Lemma~\ref{lemma6g} that
each column of $Q_\lambda T_2$ is for $u < u'$ a vector
of the form $\Psi = (\Phi, \omega \Phi)$ with $\Phi$ a solution of the
separated wave equation~(\ref{wavesep}). The same is true for $u>u'$.
Hence for $u \neq u'$, $Q_\lambda T_2$ is composed of eigenfunctions of
the Hamiltonian,
\[ \left((H-\omega) \:Q_\lambda\: T_2 \right)\!(u, \vartheta; u', \vartheta') \;=\; 0
\qquad {\mbox{if $u \neq u'$}}. \]
It remains to compute the distributional contribution to $(H-\omega) Q_\lambda
T_2$ at $u=u'$.
Since $T_2$ is continuous at $u=u'$, we only get a contribution
when both radial derivatives act on the factor $g$. According to
Lemma~\ref{lemmaspower}, the higher powers of $s$ are in $C^2$, and
thus we may replace $g$ by $s$. Applying~(\ref{hamil}),
(\ref{51a}), and Lemma~\ref{lemma61}, we obtain
\beq \label{T2eq}
\left((H-\omega) \:Q_\lambda\: T_2\right)\!(u, \vartheta; u', \vartheta')
\;=\; \frac{1}{\sigma(u, \vartheta)} \:Q_\lambda(\vartheta, \vartheta') \:
\delta(u-u')\: \left( \!\!\begin{array}{cc} 0 & 0 \\ \tau(u', \vartheta') &
\sigma(u', \vartheta') \end{array} \!\!\right) .
\eeq

We add~(\ref{T1eq}) to~(\ref{T2eq}) and carry out the sum over
$\lambda \in \sigma({\cal{A}}_\omega)$.
Since the spectral projectors $Q_\lambda$ are complete (see Lemma~\ref{lemmaangular}
{\bf{(iii)}}),
$\sum_\lambda Q_\lambda(\vartheta, \vartheta')$ gives a contribution only for
$\vartheta=\vartheta'$. We thus obtain a multiplication operator,
\[ \sum_{\lambda \in \sigma({\cal{A}}_\omega)}
(H-\omega) \:Q_\lambda \:T(\lambda) \;=\; \1 +
\left( (\beta-\omega) + \frac{\tau}{\sigma} \right)
\left(\!\!\begin{array}{cc} 0 & 0 \\
1 & 0 \end{array} \right) .  \]
Using the explicit form of the functions $\tau$, $\sigma$,
and $\beta$, one sees that the second term vanishes. Thus
\[ \sum_{\lambda \in \sigma({\cal{A}}_\omega)}
(H-\omega) \:Q_\lambda \:T(\lambda) \;=\; \1 \:. \]
Multiplying from the left by $Q_{\lambda'} S$ and using the orthogonality
of the angular spectral projectors gives the result.
\QED

\section{WKB Estimates} \label{sec7}
\setcounter{equation}{0}
In this section we shall derive estimates for the radial ODE~(\ref{5ode})
in the regime
\beq \label{regime}
|{\mbox{Re}}\, \omega| \gg 1\qquad {\mbox{and}} \qquad |{\mbox{Im}}\, \omega| \leq c\:.
\eeq
In this ``high-energy regime'', the
semi-classical WKB-solution should be a good approximation.
In order to quantify this statement rigorously, we shall
make an ansatz for $\phi$ which involves the WKB wave function and
estimate the error.

Our first lemma gives control of the sign of ${\mbox{Re}}\, \sqrt{V}$.
\begin{Lemma} \label{lemma71}
There is a constant $C$ such that for all $\omega$ with
\[ {\mbox{\rm{Im}}}\, \omega \;\neq\; 0 \;,\qquad
|{\mbox{\rm{Re}}}\, \omega| \;>\; C \]
and all $\lambda \in \sigma({\cal{A}}_\omega)$, the function
${\mbox{Re}}\, \sqrt{V}$ has no zeros.
\end{Lemma}
\Proof At a zero of ${\mbox{Re}}\, \sqrt{V}$, the function $V$ is real
and non-positive. Thus it suffices to show that the imaginary part of $V$ has
no zeros.

We first estimate the imaginary part of the angular spectrum.
For any $\lambda \in \sigma({\cal{A}}_\omega)$ we let $\Phi_\lambda$ be a
corresponding eigenvector. Then
\[ {\mbox{Im}}\, (\lambda)\: \lbra \Phi, \Phi \lket_{L^2} \;=\;
\frac{1}{2i} \left( \lbra \Phi, {\cal{A}}_\omega \Phi \lket_{L^2} -
\lbra {\cal{A}}_\omega \Phi, \Phi \lket_{L^2} \right)
\;=\; \frac{1}{2i} \lbra \Phi, ({\cal{A}}_\omega-{\cal{A}}_\omega^*) \Phi
\lket_{L^2} \;, \]
where $\lbra .,. \lket{L^2}$ is the $L^2$-scalar product on $S^2$. Hence,
according to~(\ref{angular}),
\begin{eqnarray}
|{\mbox{Im}}\, \lambda| &\leq& \frac{1}{2}
\left\| {\cal{A}}_\omega-{\cal{A}}_\omega^* \right\|
\;=\; \sup_{S^2} \left| {\mbox{Im}} \left(
\frac{1}{\sin^2 \vartheta}
(a\omega \sin^2 \vartheta + k )^{2} \right) \right| \nonumber \\
&\leq& 2 a^2\: |{\mbox{Re}}\, \omega|\: |{\mbox{Im}}\, \omega|
\:+\: \left| 2 a k \:{\mbox{Im}}\, \omega \right| \;. \label{lest}
\end{eqnarray}

The imaginary part of~(\ref{5V}) is computed to be
\beq \label{lest2}
{\mbox{Im}}\, V \;=\; -2  \left( {\mbox{Re}}\,\omega
+ \frac{ak}{r^2+a^2} \right) {\mbox{Im}}\,\omega
\:+\: \frac{\Delta}{(r^2+a^2)^2}\: {\mbox{Im}}\,\lambda
\eeq
Using~(\ref{lest}), the second summand is estimated by
\[ \left| \frac{\Delta}{(r^2+a^2)^2}\: {\mbox{Im}}\,\lambda \right|
\;\leq\; 2\: \frac{a^2 \Delta}{(r^2+a^2)^2} \left( |{\mbox{Re}}\,\omega|
+ \frac{|k|}{a} \right) |{\mbox{Im}}\,\omega| \:. \]
The factor $a^2 \Delta (r^2+a^2)^{-2}$ vanishes on the event horizon and at
infinity and is always smaller than one. Thus there is a constant $c$ with
$a^2 \Delta (r^2+a^2)^{-2} \leq c<1$. This shows that
after choosing $|{\mbox{Re}}\, \omega|$ sufficiently large, the
first summand in~(\ref{lest2}) dominates the second, and so
${\mbox{Im}}\, V$ has no zeros.
\QED
In what follows, we assume that the assumptions of the above lemma
are satisfied. We choose the sign convention for the square root such that
\begin{equation} \label{assum}
{\mbox{Re}}\, \sqrt{V(u)},\:{\mbox{Re}}\, V(u)^{\frac{1}{4}}
\;\geq\;0 \spc {\mbox{for all $u \in \R$}}.
\end{equation}
Furthermore, we shall restrict attention to $\omega$ in the range
\beq \label{wrange}
-c \;<\; {\mbox{Im}}\, \omega \;<\; 0 \;,\qquad
|{\mbox{Re}}\, \omega| \;>\; C\:,
\eeq
where $c$ is any fixed constant
and $C$ will be chosen depending on the particular application.
The next lemma, which we will need in Section~\ref{sec8}, estimates
$\sqrt{V}$ inside the ``finite box'' $[u_L, u_R]$ uniformly in $\omega$
for large ${\mbox{Re}}\, \omega$.

\begin{Lemma} \label{lemma5n}
For every angular momentum mode~$n$ and every $c, \varepsilon>0$, there
are constants $C$ and $c'$ such that for all $u \in [u_L, u_R]$ and
all $\omega$ in the range~(\ref{wrange}),
\begin{eqnarray}
|{\mbox{\rm{Re}}}\, \sqrt{V}| &\leq& c' \label{5n1} \\
|{\mbox{\rm{Im}}}\, \sqrt{V(\omega)} - {\mbox{\rm{Im}}}\, \sqrt{V({\mbox{\rm{Re}}}\,\omega)}|
&\leq& \varepsilon\:. \label{5n2}
\end{eqnarray}
\end{Lemma}
{\Proof} We set $\omega_0 = {\mbox{Re}}\, \omega$, $\lambda=\lambda(\omega_0)$
and introduce for a parameter $\tau \in [0,1]$ the potential
\begin{eqnarray*}
V_\tau &=& V(\omega_0) + \tau W \quad {\mbox{with}} \\
W &=& -2i\: {\mbox{Im}}\,\omega \left({\mbox{Re}}\, \omega + \frac{ak}{r^2+a^2} \right)
+ ({\mbox{Im}}\,\omega)^2 + (\lambda-\lambda_0)\: \frac{\Delta}{(r^2+a^2)^2} \:.
\end{eqnarray*}
Then $V_0 = V(\omega_0)$ and $V_1=V(\omega)$. The mean value theorem yields that
\begin{eqnarray}
\left| {\mbox{Re}}\, \sqrt{V(\omega)} - {\mbox{Re}}\, \sqrt{V(\omega_0)} \right|
&\leq& \sup_{\tau \in [0.1]} {\mbox{Re}} \left( \frac{W}{2 \sqrt{V_\tau}} \right) \label{75c} \\
\left| {\mbox{Im}}\, \sqrt{V(\omega)} - {\mbox{Im}}\, \sqrt{V(\omega_0)} \right|
&\leq& \sup_{\tau \in [0.1]} {\mbox{Im}} \left( \frac{W}{2 \sqrt{V_\tau}} \right) . \label{75d}
\end{eqnarray}
By choosing $C$ sufficiently large, we can clearly arrange that $V(\omega_0)<0$, and thus
$ {\mbox{Re}}\, \sqrt{V(\omega_0)}=0$. Furthermore, one sees immediately from the explicit formulas for $V$, $W$ together with the estimate for the angular eigenvalue~(\ref{ange1}) that
\begin{eqnarray*}
{\mbox{Re}}\, W &=& {\cal{O}}(|{\mbox{Re}}\, \omega|^0) \;,\qquad
{\mbox{Im}}\, W \;=\; {\cal{O}}(|{\mbox{Re}}\, \omega|^1) \\
\sqrt{V_\tau} - i {\mbox{Re}}\, \omega &=& {\cal{O}}(|{\mbox{Re}}\, \omega|^0) \:.
\end{eqnarray*}
Using this in~(\ref{75c}) and~(\ref{75d}) gives the claim.
\QED

We introduce the WKB solutions $\acute{\alpha}$ and $\grave{\alpha}$ by
\begin{equation}\label{81a}
\acute{\alpha}(u) \;=\; \acute{c}\: V^{-\frac{1}{4}}
 \: \exp\left(\int^u_0\sqrt{V} \right) ,\quad
\grave{\alpha}(u) \;=\; \grave{c}\: V^{-\frac{1}{4}}
 \: \exp\left(-\int^u_0\sqrt{V} \right) ,
\end{equation}
where~$\acute{c}$ and~$\grave{c}$ are some normalization constants.
A straightforward calculation shows that these functions satisfy the
Schr{\"o}dinger equation
\begin{equation}\label{81b}
\alpha'' \;=\; \tilde{V} \alpha \qquad\mbox{with}\qquad \tilde
V \;=\; V -\frac{1}{4}\: \frac{V''}{V} + \frac{5}{16} \left( \frac{V'}{V} \right)^2 \,.
\end{equation}
We can hope that $\acute{\alpha}$ and $\grave{\alpha}$ are
approximate solutions of the radial equation~(\ref{5ode}).
In order to estimate the error, we first write
(\ref{5ode}) as a first order system,
\begin{equation}\label{81c}
\Psi'=\left(\begin{array}{cc}0 &1\\ V&0\end{array}\right)
\Psi\spc \mbox{with}\spc \Psi=\left(\begin{array}{c}\phi\\\phi'\end{array}\right) .
\end{equation}
Next we make for $\Psi$ the ansatz
\begin{equation}\label{81d}
\Psi = A\Phi\spc \mbox{with}\spc A=\left(\begin{array}{cc}
\acute{\alpha} & \grave{\alpha} \\ \acute{\alpha}' & \grave{\alpha}'
\end{array}\right)
\end{equation}
and $\Phi$ a $2$-component complex function. $A$ is the fundamental
matrix of the ODE (\ref{81b}) and thus
\begin{equation}\label{84a}
A'=\left(\begin{array}{cc} 0 &1\\\tilde{V} &0\end{array}\right) A \;.
\end{equation}
Differentiating through the ansatz for (\ref{81d}) and using
(\ref{81c}, \ref{84a}), we obtain that
\begin{equation}\label{81e}
A\:\Phi' = \left(\begin{array}{cc} 0 &0\\ V-\tilde{V}
    &0\end{array}\right)
A\:\Phi\;.
\end{equation}
The determinant of $A$ is a Wronskian and thus constant. A short
computation using (\ref{81a}) shows that $\det A = -2 \acute{c} \grave{c}.$
Hence we can easily compute the inverse of~$A$ by Cramer's rule,
$$A^{-1} = -\frac{1}{2 \acute{c} \grave{c}}\left(\begin{array}{rr} \grave{\alpha}' & -\grave{\alpha} \\
-\acute{\alpha}' & \acute{\alpha} \end{array}\right)\,,$$
and multiplying (\ref{81e}) by $A^{-1}$ gives
$$\Phi'=-\frac{1}{2 \acute{c} \grave{c}}\: (V-\tilde{V}) \left(\begin{array}{cc}
-\grave{\alpha} \acute{\alpha}
  &-\grave{\alpha}^2 \\ \acute{\alpha}^2
& \acute{\alpha} \grave{\alpha} \end{array}\right) \Phi\,.$$
Finally, we put in the explicit formulas (\ref{81b}) and (\ref{81a})
to obtain the equation
\begin{equation}\label{81f}
\Phi'=W\left(\begin{array}{rc}-1 &-f^{-1}\\f&1\end{array}\right) \Phi\;,
\end{equation}
where $W$ and $f$ are the functions
\beq \label{Wf}
 W \;=\; \frac{1}{8 \acute{c} \grave{c}}\left( \frac{V''}{V^{\frac{3}{2}}} 
 \:-\: \frac{5}{4}\: \frac{V'^2}{V^\frac{5}{2}}
 \right)\;,\qquad
f \;=\; \exp\left(2\int^u_0\sqrt{V}\right)\,.
\eeq

We shall now derive an estimate for the solutions of the ODE~(\ref{81f}).
The main difficulty is that when the function $f$
is very large or close to zero, the matrix in~(\ref{81f}) has large
norm, making it impossible to use simple Gronwall estimates.
Instead, we can use that according to~(\ref{assum}), the function
\[ |f| \;=\; \exp\left(2\int^u_0 {\mbox{Re}}\, \sqrt{V} \right) \]
is monotone.

\begin{Thm} \label{lemma72}
Assume that the potential $V$ in the Schr{\"o}dinger
equation~(\ref{5ode}) satisfies the conditions~(\ref{assum})
and that the function $W$ defined by~(\ref{Wf}) is in~$L^1(\R)$.
Then there is a solution $\Phi$ of the system of ODEs~(\ref{81f})
with boundary conditions
\[ \lim_{u \to -\infty} \Phi(u) \;=\; \left( \!\!\begin{array}{c} 1 \\ 0 \end{array}
\!\!\right). \]
This solution satisfies for all $u \in \R$ the bounds
\begin{eqnarray*}
\left|\Phi_1(u) - \exp\left(-\int_{-\infty}^u W \right) \right| &\leq& e^{4 \|W\|_1}\: \|W\|_1 \\
\left|\Phi_2(u) \right| &\leq& e^{4 \|W\|_1}\: \|W\|_1 \;|f(u)|\:.
\end{eqnarray*}
\end{Thm}
\Proof We set $\rho=|f|$ and introduce new functions~$a$ and~$b$ by
\begin{equation} \label{bdef}
a=\Phi_1 \qquad\mbox{and}\qquad b=\frac{\Phi_2}{\rho}\,.
\end{equation}
According to~(\ref{81f}), they satisfy the following ODEs,
\begin{eqnarray*}
a'&=&-Wa-W\frac{\rho}{f} b\\
b'+\frac{\rho'}{\rho}b &=& W\frac{f}{\rho}a+Wb\,.
\end{eqnarray*}
This gives rise to the following differential inequalities,
\begin{eqnarray*}
|a|' &=& \frac{d}{du} \sqrt{\overline{a} a} \;=\;
\frac{1}{|a|}\: {\mbox{Re}} \left( \overline{a}\:a' \right)
\;\leq\; -|a|\:{\mbox{Re}}\, W \:+\: |b| \left|W \:\frac{\rho}{f} \right| \\
|b|' &=& \frac{1}{|b|}\: {\mbox{Re}} \left( \overline{b}\:b' \right)
\;\leq\; -|b|\: \frac{\rho'}{\rho} \:+\: |a| \left|W\: \frac{f}{\rho} \right|
\:+\: |b|\: {\mbox{Re}}\, W \;.
\end{eqnarray*}
Using that $\rho$ is monotone and that $|f|=\rho$, we obtain the simple
inequality
\[ \left( |a|+|b| \right)' \;\leq\; 2 |W|\: (|a|+|b|) \;. \]
Integrating this inequality from $v$ to $u$, $-\infty<v<u<\infty$, gives
the ``Gronwall estimate''
\begin{equation}\label{81i}
(|a|+|b|)(u) \;\leq\; (|a|+|b|)(v)\:
\exp \left( 2 \int_{v}^u |W| \right) \;\leq\; (|a|+|b|)(v)\:e^{2\|W\|_1} \:.
\end{equation}

We now let $\Phi^{(v)}$ be the solution of~(\ref{81f}) with boundary
conditions $\Phi^{(v)}(v)=(1,0)$. In order to estimate~$\Phi^{(v)}$, we
rewrite~(\ref{81f}) as
\begin{eqnarray*}
\left( e^{\int_{v}^u W} \:\Phi_1^{(v)}(u) \right)' &=& -W\:e^{\int_{v}^u W}\: \frac{\Phi_2^{(v)}}{f} \\
\left( e^{-\int_{v}^u W} \:\Phi_2^{(v)}(u) \right)' &=& W\:e^{-\int_{v}^u W}\: f\:\Phi_1^{(v)} \:.
\end{eqnarray*}
We integrate and use~(\ref{81i}) to obtain
the inequalities
\begin{eqnarray}
\left| e^{\int_{v}^u W} \:\Phi_1^{(v)}(u) - 1 \right| &\leq& e^{3\|W\|_1} \int_{v}^u |W| \label{est1} \\
\left| e^{-\int_{v}^u W} \:\Phi_2^{(v)}(u) \right| &\leq& e^{3\|W\|_1} \int_{v}^u \rho\:W \;\leq\;
e^{3\|W\|_1}\:\rho(u) \int_{v}^u |W| \;, \label{est2}
\end{eqnarray}
where in the last step we used the monotonicity of $\rho$.

The inequalities~(\ref{est1}) and~(\ref{est2}) yield that for every
$\varepsilon>0$, there is a $\tilde{u}$ such that for all $v,v'<\tilde{u}$,
the exponential on the left side of~(\ref{est1}) and~(\ref{est2}) are arbitrarily
close to one, and the integrals on the right can be made arbitrarily small.
Thus $|(\Phi^{(v)}-\Phi^{(v')})(\tilde{u})|<\varepsilon$. Due to the factor $\rho(u)$ on the right
of~(\ref{est2}), we even know  that $|(b^{(v)}-b^{(v')})(\tilde{u})|<\varepsilon$
(with $b$ according to~(\ref{bdef})).
Since~(\ref{81f}) is linear, $\Phi^{(v)}-\Phi^{(v')}$ is also a solution.
Applying~(\ref{81i}) for this solution and choosing $v=\tilde{u}$, we obtain
that for all $u>\tilde{u}$,
$|(\Phi^{(v)}-\Phi^{(v')})(u)|<c \varepsilon$ with a constant $c$ being independent
of $\tilde{u}$. This shows that $\Phi^{(v)}(u)$ converges as $v \to -\infty$,
and that the above estimates are still true for $v=-\infty$.

The theorem now follows from~(\ref{est1}) and~(\ref{est2}) if we
set $v=-\infty$ and pull out a factor of
$e^{\int_{-\infty}^u W}$ and $e^{-\int_{-\infty}^u W}$, respectively.
\QED

The above theorem has two immediate consequences: First, it yields
the existence of solutions $\acute{\phi}$ and $\grave{\phi}$
which decay exponentially at minus and plus infinity, respectively.
Second, it gives very good control of the global behavior of these
solutions if $|{\mbox{Re}}\, \omega|$ is large.
\begin{Corollary} \label{cor74}
For every angular momentum mode~$n$ and every
$\omega$ with ${\mbox{\rm{Im}}} \,\omega <0$, there are solutions
$\acute{\phi}$ and $\grave{\phi}$ of the Schr{\"o}dinger
equation~(\ref{5ode}) which satisfy the boundary conditions
\[ \lim_{u \to -\infty} \left| e^{-i\omega u} \:\acute{\phi}(u) \right| \;=\; 1 \;=\;
\lim_{u \to \infty} \left| e^{i\omega u} \:\grave{\phi}(u) \right| \:. \]
\end{Corollary}
{\Proof}
It suffices to construct $\acute{\phi}$, because~$\grave{\phi}$ is obtained
in exactly the same way if one considers the ODEs backwards in $u$
(i.e.\ after transforming the radial variable according to $u \to -u$). We choose $\Phi$ as in
Theorem~\ref{lemma72} and let
$\acute{\phi}=\phi$ be the corresponding solution of the Schr{\"o}dinger
equation given by~(\ref{81d}) and~(\ref{81c}).
Note that the corollary only makes a statement on the asymptotic behavior
of $\acute{\phi}$ as $u \to -\infty$, and thus the
behavior of $\acute{\phi}$ on any interval $[u_0, \infty)$, $u_0<0$
is irrelevant. Thus we may freely modify the potential $V$ on
any such interval. In particular, we can
change the potential $V$ on $[u_0, \infty)$ such that it is constant
for large $u$. For any $\varepsilon>0$, we choose $u_0$ so small and
modify $V_{|[u_0, \infty)}$ such that ${\mbox{Re}}\, \sqrt{V} \geq 0$
and~$\|W\|_1< \varepsilon/3$ (this is possible because $V''$ decays for large
$|u|$ at least at the rate $\sim |u|^{-3}$). Then Theorem~\ref{lemma72} applies,
and we obtain that
\[ |\Phi_1-1| \;\leq\; \varepsilon \;,\qquad |\Phi_2| \;\leq\; \varepsilon\: |f| \:. \]
Using these bounds in~(\ref{81d}), one sees that $|\phi/ \acute{\alpha} - 1|
<\varepsilon$ and thus, after choosing the normalization constants
$\acute{c}$ and $\grave{c}$ in~(\ref{81a}) appropriately,
\[ \limsup_{u \to -\infty} |e^{-i \omega u}\: \acute{\phi}| \;\leq\; 1+\varepsilon \quad {\mbox{and}} \quad
\liminf_{u \to -\infty} |e^{-i \omega u}\: \acute{\phi}| \;\geq\; 1-\varepsilon \:. \]
Since $\varepsilon$ is arbitrary, the result follows.
\QED

\begin{Prp} \label{prp74}
For every $n$ and $c, \varepsilon>0$,
there is a constant~$C>0$ such that for all $\omega$ in the range~(\ref{wrange}),
the solutions $\acute{\phi}$ and $\grave{\phi}$ of Corollary~\ref{cor74}
are close to the (suitably normalized) WKB wave functions
$\acute{\alpha}$ and $\grave{\alpha}$,
(\ref{81a}), in the sense that for all $u \in \R$,
\[ \left|\frac{\acute{\phi}}{\acute{\alpha}} - 1 \right|
+\left|\frac{\acute{\phi'}}{\acute{\alpha}'} - 1 \right| \;\leq\; \varepsilon
\qquad {\mbox{and}} \qquad
\left|\frac{\grave{\phi}}{\grave{\alpha}} - 1 \right|
+ \left|\frac{\grave{\phi'}}{\grave{\alpha}'} - 1 \right| \;\leq\; \varepsilon \:. \]
\end{Prp}
The reason why we need to choose $C$ large is that the functions
$V'/|V|^{3/2}$ and $W$ must be sufficiently small.
More specifically, one can choose $C$
such that
\[ \|W\|_1 \;\leq\; \frac{\varepsilon}{3}
\qquad {\mbox{and}}\qquad \frac{|V'|}{|V|^\frac{3}{2}} \;\leq\;
\frac{1}{3} \:. \]

\noindent
{\em{Proof of Proposition~\ref{prp74}.}}
Using~(\ref{ange1}) in~(\ref{5V}), one sees that in the strip $-2c < {\mbox{Im}}\, \omega < 0$, the potential $V$ satisfies the bound
\[ |V(u) + \omega^2| \;\leq\; c_1 (1+|\omega|) \;. \]
On the other hand, differentiating~(\ref{5V}) and using~(\ref{51a})
and~(\ref{ange1}), one sees that in this strip,
\[ |V'(u)| + |V''(u)| \;\leq\; (1+|\omega|)\: g(u) \;, \]
where $g$ is a function which decays for large $|u|$ at least at the rate~$\sim u^{-2}$. Putting these estimates
for $V$ and $V''$ into~(\ref{Wf}), one sees that by choosing $C$
sufficiently large, we can arrange that for all $\omega$ in the
range~(\ref{wrange}), $\|W\|_1 \leq \varepsilon/3$.
Theorem~\ref{lemma72} yields that
\beq \label{Pes}
|\Phi_1-1| \;\leq\; \varepsilon \;,\qquad
|\Phi_2| \;\leq\; \varepsilon\: |f| \:.
\eeq
Dividing the first row in~(\ref{81d}) by $\acute{\alpha}$, we obtain
the identity
\[ \frac{\acute{\phi}}{\acute{\alpha}} \;=\; \Phi_1 +
\frac{\grave{\alpha}}{\acute{\alpha}}\: \Phi_2\:, \]
and using~(\ref{Pes}) gives
\[ \left| \frac{\acute{\phi}}{\acute{\alpha}} - 1 \right|
\;\leq\; \varepsilon \left(1 + |f| \left|
\frac{\grave{\alpha}}{\acute{\alpha}} \right|
\right) . \]
>From the second row in~(\ref{81d}) we obtain similarly,
\[ \left| \frac{\acute{\phi}'}{\acute{\alpha}'} - 1 \right|
\;\leq\; \varepsilon \left(1 + |f| \left| \frac{\grave{\alpha}'}{\acute{\alpha}'} \right|
\right) . \]
Finally, we apply the elementary estimates for the WKB wave functions
\[ \left| \frac{\grave{\alpha}}{\acute{\alpha}} \right| \;=\; \frac{1}{|f|} \;,\spc
\left| \frac{\grave{\alpha}'}{\acute{\alpha}'} \right| \;=\; \frac{1}{|f|}
\left|\frac{V'\:V^{-\frac{5}{4}} + V^{\frac{1}{4}}}{V'\:V^{-\frac{5}{4}} - V^{\frac{1}{4}}}
\right| \;\leq\; \frac{2}{|f|} \:, \]
where in the last step we applied the above bounds for $V'$ and $V$
and possibly increased $C$.

The solution $\grave{\phi}$ is obtained similarly if one
considers the Schr{\"o}dinger equation~(\ref{5ode}) backwards in $u$
and repeats the above arguments.
\QED

The next two propositions give estimates for composite expressions.
\begin{Prp} \label{prp75}
Under the assumptions of Proposition~\ref{prp74},
\[ \left| \frac{w(\acute{\phi}, \grave{\phi})}{w(\acute{\alpha}, \grave{\alpha})}
-1 \right| \;\leq\; 4 \varepsilon\:. \]
\end{Prp}
{\Proof} Rewriting the Wronskian as
\[ w(\acute{\phi}, \grave{\phi}) \;=\; \acute{\phi}' \: \grave{\phi} -
\acute{\phi}\: \grave{\phi}' \;=\;
\acute{\alpha}'\: \grave{\alpha}\; \frac{\acute{\phi}'}{\acute{\alpha}'}
\: \frac{\grave{\phi}}{\grave{\alpha}}
- \acute{\alpha}\: \grave{\alpha}'\; \frac{\acute{\phi}}{\acute{\alpha}}
\: \frac{\grave{\phi}'}{\grave{\alpha}'} \:, \]
we can put in the estimate of Proposition~\ref{prp74} to obtain
\begin{equation} \label{75a}
\left| w(\acute{\phi}, \grave{\phi}) - w(\acute{\alpha}, \grave{\alpha}) \right|
\;\leq\; 4 \varepsilon \left( |\acute{\alpha}'\: \grave{\alpha}| +
|\acute{\alpha}\: \grave{\alpha}'| \right) .
\end{equation}
Furthermore, a short explicit calculation using~(\ref{81a}) shows that
\begin{eqnarray}
w(\acute{\alpha}, \grave{\alpha}) &=& 2 \sqrt{V}\: \acute{\alpha} \grave{\alpha} \label{75x} \\
(\acute{\alpha}\: \grave{\alpha})' &=& -\frac{1}{2}\: \frac{V'}{V}\: \acute{\alpha} \grave{\alpha}
\;=\; -\frac{1}{4}\: V'\: V^{-\frac{3}{2}} \: w(\acute{\alpha}, \grave{\alpha})
\end{eqnarray}
and thus
\begin{eqnarray*}
\acute{\alpha}' \: \grave{\alpha} &=& \frac{1}{2} \left( w(\acute{\alpha}, \grave{\alpha})
+ (\acute{\alpha}\: \grave{\alpha})' \right) \;=\;
\frac{1}{2}\: w(\acute{\alpha}, \grave{\alpha}) \left( 1 - \frac{1}{4}\: V'\: V^{-\frac{3}{2}} \right) \\
\acute{\alpha} \: \grave{\alpha}' &=& -\frac{1}{2} \left( w(\acute{\alpha}, \grave{\alpha})
- (\acute{\alpha}\: \grave{\alpha})' \right) \;=\;
-\frac{1}{2}\: w(\acute{\alpha}, \grave{\alpha}) \left( 1 + \frac{1}{4}\: V'\: V^{-\frac{3}{2}} \right) .
\end{eqnarray*}
Substituting these relations into~(\ref{75a}) gives
\[ \left| \frac{w(\acute{\phi}, \grave{\phi})}{w(\acute{\alpha}, \grave{\alpha})} - 1 \right|
\;\leq\; 4 \varepsilon \left( 1 + \frac{1}{4} \left| V'\: V^{-\frac{3}{2}} \right| \right) . \]
Here the left side only involves Wronskians and is thus independent of $u$. Hence we may
on the right side take the limit $u \to \infty$. This gives the result.
\QED

For $u_L<u_R$ we set
\beq \label{7nota}
\left. \begin{array}{rcl}
\phi_{[u_L, u_R]} &=& \acute{\phi}(u_L)\: \grave{\phi}(u_R) -
\acute{\phi}(u_R)\: \grave{\phi}(u_L) \\[.5em]
\alpha_{[u_L, u_R]} &=& \acute{\alpha}(u_L)\: \grave{\alpha}(u_R) -
\acute{\alpha}(u_R)\: \grave{\alpha}(u_L)\:.
\end{array} \qquad \right\}
\eeq

\begin{Prp} \label{prp76}
Under the assumptions of Proposition~\ref{prp74},
\[ \left| \frac{\phi_{[u_L, u_R]}}{\alpha_{[u_L, u_R]}} - 1 \right|
\;\leq\; 8 \varepsilon \:\exp \left( 2\int_{u_L}^{u_R} {\mbox{Re}}\, \sqrt{V} \right)\:
 \left| \sin \left( 2 \int_{u_L}^{u_R} {\mbox{\rm{Im}}}\, \sqrt{V} \right)
\right|^{-1} . \]
\end{Prp}
{\Proof} Rewriting $\phi_{[u_L, u_R]}$ as
\[ \phi_{[u_L, u_R]} \;=\; \acute{\alpha}(u_L)\: \grave{\alpha}(u_R) \:
\frac{\acute{\phi}(u_L)}{\acute{\alpha}(u_L)}\: \frac{\grave{\phi}(u_R)}{\grave{\alpha}(u_R)}
\:-\: \acute{\alpha}(u_R)\: \grave{\alpha}(u_L) \:
\frac{\acute{\phi}(u_R)}{\acute{\alpha}(u_R)}\: \frac{\grave{\phi}(u_L)}{\grave{\alpha}(u_L)} \:, \]
Proposition~\ref{prp74} yields that
\beq \label{75b}
\left| \phi_{[u_L, u_R]} - \alpha_{[u_L, u_R]} \right| \;\leq\;
4 \varepsilon \left( |\acute{\alpha}(u_L)\: \grave{\alpha}(u_R)| +
|\acute{\alpha}(u_R)\: \grave{\alpha}(u_L)| \right)
\eeq
Furthermore, it is obvious from~(\ref{81a}) that
\[ \acute{\alpha}(u_L)\: \grave{\alpha}(u_R) \;=\;
\acute{\alpha}(u_R)\: \grave{\alpha}(u_L)\: \exp \left( -2\int_{u_L}^{u_R} \sqrt{V} \right) \]
and thus from~(\ref{7nota}),
\begin{eqnarray}
\alpha_{[u_L, u_R]} &=& \acute{\alpha}(u_R)\: \grave{\alpha}(u_L) \left(
\exp \left( -2\int_{u_L}^{u_R} \sqrt{V} \right) - 1 \right) \nonumber \\
&=& \acute{\alpha}(u_L)\: \grave{\alpha}(u_R) \left(
1 - \exp \left( 2\int_{u_L}^{u_R} \sqrt{V} \right) \right) . \label{7aLR}
\end{eqnarray}
Dividing~(\ref{75b}) by $\alpha_{[u_L, u_R]}$ and putting in the last identities, we
obtain
\[ \left| \frac{\phi_{[u_L, u_R]}}{\alpha_{[u_L, u_R]}} - 1 \right|
\;\leq\; 4 \varepsilon\: \frac{1 + \left| \exp \left( -2\int_{u_L}^{u_R} \sqrt{V} \right) \right|}
{\left| \exp \left( -2\int_{u_L}^{u_R} \sqrt{V} \right) - 1 \right|}
\;\leq\;  \frac{8 \varepsilon}
{\left| \exp \left( -2\int_{u_L}^{u_R} \sqrt{V} \right) - 1 \right|} , \]
where in the last step we used that ${\mbox{Re}}\, \sqrt{V} \geq 0$.
We finally estimate the obtained denominator from above,
\begin{eqnarray*}
\lefteqn{ \left| \exp \left( -2\int_{u_L}^{u_R} \sqrt{V} \right) - 1 \right|
\;\geq\; \left| {\mbox{Im}} \exp \left( -2\int_{u_L}^{u_R} \sqrt{V} \right) \right| } \\
&=& \exp \left( -2\int_{u_L}^{u_R} {\mbox{Re}}\, \sqrt{V} \right)\:
\left| \sin \left( 2\int_{u_L}^{u_R} {\mbox{Im}}\, \sqrt{V} \right) \right| .
\end{eqnarray*}
\QED

\section{Contour Deformations} \label{sec8}
\begin{figure}[tb]
\begin{center}
\input{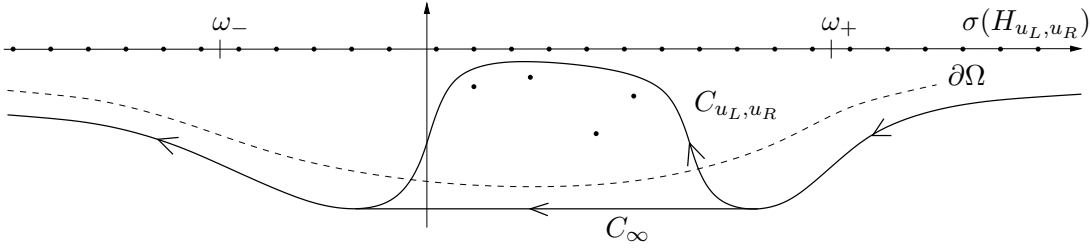}
\caption{Unbounded contour representation of $E_\C$}
\label{fig1}
\end{center}
\end{figure}

\begin{figure}[tb]
\begin{center}
\input{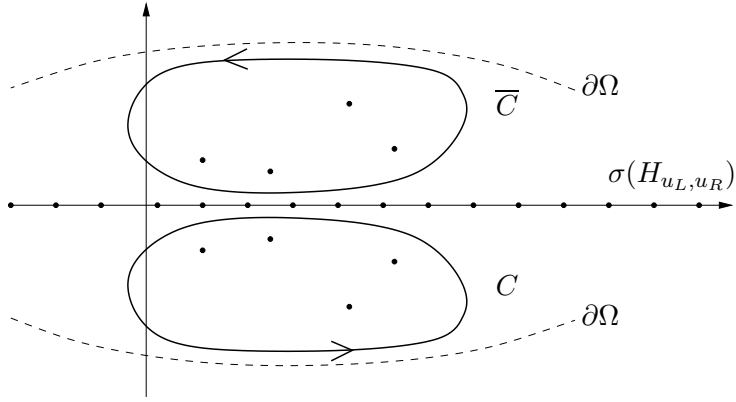}
\caption{Closed contour representation of $E_\C$}
\label{fig2}
\end{center}
\end{figure}

\setcounter{equation}{0}
In this section we shall use contour integral methods to prove
the main theorem. Recall that in Section~\ref{sec3}, we showed
that the Hamiltonian $H_{u_L, u_R}$ in finite
volume is a selfadjoint operator on the Pontrjagin space
${\cal{P}}_{u_L, u_R}$. It has a purely discrete spectrum, and for
each $\omega \in \sigma(H_{u_L, u_R})$, the projector $E_\omega$ onto
the corresponding invariant subspace can be expressed
as the contour integral
\[ E_\omega \;=\; -\frac{1}{2 \pi i} \oint_{|\omega'-\omega| < \varepsilon}
S_{u_L, u_R}(\omega')\: d\omega' \:, \]
where $\varepsilon$ is to be chosen so small that $B_\varepsilon(\omega)$
contains no other points of the spectrum.
The theory of Pontrjagin spaces also yields that $\sigma(H_{u_L, u_R})$
will in general involve a finite number of non-real spectral points, which
lie symmetrically around the real axis. We let $E_\C$ be the projector
onto the invariant subspace corresponding to all non-real spectral
points,
\beq \label{EC}
E_\C \;:=\; \sum_{\omega \in \sigma(H_{u_L, u_R}) \setminus \R}
E_\omega \:.
\eeq
Our first lemma represents $E_\C$ as a Cauchy integral over
an unbounded contour. More precisely, we choose a contour~$C_{u_L, u_R}$
in the lower half plane which joins the points $+\infty$ with $-\infty$ and
encloses the spectrum in the lower half plane from above.
Furthermore, if ${\mbox{Re}}\, \omega$ is outside the
finite interval $[\omega_-, \omega_+]$, $\omega$ should
be in the open set $\Omega$ (see~(\ref{Odef}))
and should approach the real axis as $|{\mbox{Im}}\, \omega|
\sim -|{\mbox{Re}}\, \omega|^{-1}$
(see Figure~\ref{fig1}).
\begin{Lemma} \label{lemma91}
The spectral projector corresponding to the non-real
spectrum~(\ref{EC}) has the representation
\beq \label{crep1}
E_\C \;=\; \frac{1}{\pi}\: \lim_{L \to \infty}
{\mbox{\rm{Im}}} \int_{C_{u_L, u_R}} \frac{L^3}{(L+i \omega)^3}\:
S_{u_L, u_R}(\omega)\:d\omega\:.
\eeq
\end{Lemma}
{\Proof} The Cauchy integral formula yields that
\[ E_\C \;=\; -\frac{1}{2 \pi i} \oint_C S_{u_L, u_R}(\omega)\: d\omega
\:+\: \frac{1}{2 \pi i} \oint_{\overline{C}}
S_{u_L, u_R}(\omega)\: d\omega
\;=\; \frac{1}{\pi} \:{\mbox{Im}} \oint_C S_{u_L, u_R}(\omega)\:
d\omega \:, \]
where $C$ is a closed contour which encloses the spectrum in the lower
half plane (see Figure~\ref{fig2}).
The dominated convergence theorem allows us to insert a factor
$L^3/(L+i \omega)^3$,
\[ E_\C \;=\; \frac{1}{\pi} \:\lim_{L \to \infty}\:
{\mbox{Im}} \oint_C \frac{L^3}{(L+i \omega)^3}\:
S_{u_L, u_R}(\omega)\: d\omega \:. \]
The function $L^3/(L+i \omega)^3$ has no poles in
the lower half plane and decays cubically for large $|\omega|$.
Furthermore, according to~(\ref{Sbd2}), the resolvent grows
at most linearly for large $|\omega|$.
This allows us to deform the contour in such a way that
$C$ is closed in the lower half plane on larger
and larger circles $|\omega|=R$. In the limit $R \to \infty$
the contribution along the circle tends to zero. Thus we end up with the integral along the contour~$C_{u_L, u_R}$.
\QED

Our next goal is to get rid of the ``convergence generating factor''
$L^3/(L+i \omega)^3$ in~(\ref{crep1}). We shall use the fact that
when we take the difference $S_{u_L, u_R} - S_\infty$ and evaluate
it with a test function, the resulting expression has much better
decay properties at infinity (see Lemma~\ref{lemma54}).
We choose a contour $C_\infty$ which coincides with $C_{u_L, u_R}$
if ${\mbox{Re}}\, \omega \not \in [\omega_-, \omega_+]$ and
always stays inside $\Omega$ (see Figure~\ref{fig1}).
\begin{Lemma} \label{lemma92}
For every $\Psi \in C^\infty_0((u_L, u_R) \times S^2)^2$,
\beq
\bra \Psi, E_\C \Psi \ket \;=\; \label{crep2}
\frac{1}{\pi} \: {\mbox{\rm{Im}}}
\left( \int_{C_{u_L, u_R}} \bra \Psi, S_{u_L, u_R} \Psi \ket\: d\omega
\:-\: \int_{C_\infty} \bra \Psi, S_\infty \Psi \ket \: d\omega \right) .
\eeq
Furthermore,
\begin{eqnarray}  \label{crep3}
\bra \Psi, E_\C \Psi \ket &=& \sum_{n \in \sN} I_n \qquad {\mbox{with}} \\
I_n &=& -\frac{1}{2\pi i}
\left( \int_{C_{u_L, u_R}} \bra \Psi, Q_n \:S_{u_L, u_R} \Psi \ket\:
d\omega \:-\: \int_{C_\infty} \bra \Psi, Q_n S_\infty \Psi \ket \: d\omega
\right) \nonumber \\
&&+\frac{1}{2\pi i}
\left( \int_{\overline{C_{u_L, u_R}}} \bra \Psi,
Q_n \:S_{u_L, u_R} \Psi \ket\:
d\omega \:-\: \int_{\overline{C_\infty}}
\bra \Psi, Q_n S_\infty \Psi \ket \: d\omega \right). \quad\;\;\;\; \label{Indef}
\end{eqnarray}
The series in~(\ref{crep3}) converges absolutely.
\end{Lemma}
We point out that the above integrals are merely a convenient
notation and are to be given a rigorous meaning as follows.
We formally rewrite the integrals in~(\ref{crep2}) (and similarly
in~(\ref{Indef})) as
\beq \label{rewrite}
\int_{C_\infty} \bra \Psi, (S_{u_L, u_R}-S_\infty) \Psi \ket
\:d\omega \:+\: \left( \int_{C_{u_L, u_R}} \!\!- \int_{C_\infty} \right)
\bra \Psi, S_{u_L, u_R} \Psi \ket \:d\omega \:.
\eeq
Now the first summand is well-defined according to Lemma~\ref{lemma54}.
In the second summand, the integrals combine to an
integral over a bounded contour, and this is clearly well-defined because
the contour does not intersect the spectrum of $H_{u_L, u_R}$.

Note that in~(\ref{Indef}) we cannot combine the integrals over
$C_{u_L, u_R}$ and $\overline{C_{u_L, u_R}}$
(and similarly over $C_\infty$ and $\overline{C_\infty}$)
to the imaginary part of one contour integral
because $Q_n$ in general does not commute with $S_{u_L, u_R}$, and so the
integrands in~(\ref{Indef}) need not be real. For notational convenience,
we abbreviate the second line in~(\ref{Indef}) by ``$-{\mbox{ccc}}$''
(for ``complex conjugated contours'').
 \\[.5em]
{\em{Proof of Lemma~\ref{lemma92}. }} According to Corollary~\ref{cor53},
the resolvent $S_\infty(\omega)$ is holomorphic for $\omega in \Omega$
and grows at most linearly in $|\omega|$. Thus for all $L>0$,
\[ \int_{C_\infty} \frac{L^3}{(L+i \omega)^3}\: S_\infty(\omega)\:d\omega
\;=\; 0 \:. \]
Combining this identity with~(\ref{crep1}), we obtain the representation
\begin{eqnarray*}
\lefteqn{ \bra \Psi, E_\C \Psi \ket \;=\;
\frac{1}{\pi}\: \lim_{L \to \infty} } \\
&& \times \:{\mbox{\rm{Im}}} \left(
\int_{C_{u_L, u_R}} \frac{L^3}{(L+i \omega)^3}\:
\bra \Psi, S_{u_L, u_R} \Psi \ket \:d\omega
\:-\: \int_{C_\infty} \frac{L^3}{(L+i \omega)^3}\:
\bra \Psi, S_\infty \Psi \ket \:d\omega \right) .
\end{eqnarray*}
We now rewrite the integrals according to~(\ref{rewrite}).
If replace the contour $C_{u_L, u_R}$ by
$C_\infty$, the integrands combine, and we obtain the expression
\[ \frac{1}{\pi}\: \lim_{L \to \infty}
\:{\mbox{\rm{Im}}} \int_{C_\infty} \frac{L^3}{(L+i \omega)^3}\:
\bra \Psi, (S_{u_L, u_R}-S_\infty) \Psi \ket \:d\omega \:. \]
The estimate~(\ref{res1}) allows us to apply Lebesgue's dominated
converge theorem and to take the limit $L \to \infty$ inside the integrand. The error we made when replacing $C_{u_L, u_R}$ by
$C_\infty$ is
\[ \frac{1}{\pi}\: \lim_{L \to \infty}
\:{\mbox{\rm{Im}}} \left\{ \left(
\int_{C_{u_L, u_R}} \!\!- \int_{C_\infty} \right)
\frac{L^3}{(L+i \omega)^3}\: \bra \Psi, S_{u_L, u_R} \Psi \ket \:d\omega
\right\} . \]
Now the contour is bounded, and since
the factor~$\bra \Psi, S_{u_L, u_R} \Psi \ket$ is bounded,
we can again apply Lebesgue's dominated convergence theorem
to take the limit $L \to \infty$
inside the integrand. This gives~(\ref{crep2}).

Note that our contours were chosen such that the condition~(\ref{ocond})
is satisfied for a suitable constant $c>0$, and so Lemma~\ref{lemmaangular}
applies.
Using completeness of the $(Q_n)_{n \in \sN}$ (see Lemma~\ref{lemmaangular} {\bf{(iii)}}),
it immediately follows from~(\ref{crep2}) that
\begin{eqnarray*}
\lefteqn{ \bra \Psi, E_\C \Psi \ket } \\
&=&
-\frac{1}{2 \pi i} \left( \int_{C_{u_L, u_R}} \sum_{n \in \sN}
\bra \Psi, Q_n S_{u_L, u_R} \Psi \ket\: d\omega
\:-\: \int_{C_\infty} \sum_{n \in \sN} \bra \Psi, Q_n
S_\infty \Psi \ket \: d\omega \right) - {\mbox{ccc}}\: .
\end{eqnarray*}
Again replacing the contour $C_{u_L, u_R}$ by
$C_\infty$, we obtain the expression
\[ -\frac{1}{2\pi i}
\int_{C_\infty} \sum_{n \in \sN}
\bra \Psi, Q_n \:(S_{u_L, u_R}-S_\infty) \Psi \ket\: d\omega\:
- {\mbox{ccc}}\: . \]
According to~(\ref{res2}), the summands decay faster than any polynomial
in $\lambda_n$.
Applying the angular estimates~(\ref{ang11}) and~(\ref{Qnb}), we conclude that the sum
over $n$ converges absolutely, uniformly in $\omega \in C_\infty$.
Thus the dominated convergence theorem allows us to commute summation and
integration, and the series converges absolutely. It remains to
consider the expression
\[ -\frac{1}{2 \pi i} \left( \int_{C_{u_L, u_R}} - \int_{C_\infty} \right) \sum_{n \in \sN}
\bra \Psi, Q_n S_{u_L, u_R} \Psi \ket\: d\omega \: - {\mbox{ccc}}\:. \]
Now the contours are compact, and thus the absolute convergence
of the $n$-series is uniform
on the contour. Hence we can again apply Lebesgue's dominated
convergence theorem to interchange the summation with the integration.
\QED

We shall now deform the contours $C_{u_L, u_R}$ and
$C_\infty$ and analyze the resulting integrals. Our aim is to move
the contours onto the real axis such that they reduce to an
$\omega$-integral over the real line.
It is a major advantage of~(\ref{crep3}) that
the series stands in front of the integrals, because this allows us to
deform the contours in each summand $I_n$ separately. Moreover, since our
contour deformations will keep the values of the integrals unchanged,
Lemma~\ref{lemma92} guarantees that the series over $n$ will converge
absolutely. Thus we may in what follows restrict attention to fixed $n$.

For given $n$, we know from Section~\ref{sec3} that
the function $\bra \Psi, Q_n S_{u_L, u_R} \Psi \ket$ is
meromorphic, and all poles are points of~$\sigma(H_{u_L, u_R})$.
For the integrals over $C_\infty$ in~(\ref{Indef}),
we cannot use abstract arguments
because we have hardly any information on the spectrum of $H_\infty$
(we only know from Lemma~\ref{lemma52imp} that the spectrum lies
outside the set $\Omega$, (\ref{Odef}), but it may be
continuous and complex). But from the separation of the resolvent we know
that the operator $Q_\lambda S_\infty$ is well-defined and bounded
unless the Wronskian $w(\acute{\phi}, \grave{\phi})$ vanishes
(see Proposition~\ref{thm63} and~(\ref{sdef})). If this Wronskian
were zero and ${\mbox{Im}}\, \omega<0$, this would give rise to a solution
$\phi$ of the reduced wave equation which decays exponentially as
$u \to \pm \infty$. Such ``unstable modes'' were ruled out
by Whiting~\cite{W}. We conclude that $\bra \Psi, Q_n S_\infty \Psi \ket$
is analytic in the whole lower half plane $\{{\mbox{Im}}\, \omega <0\}$.

\begin{figure}[htbp]
\begin{center}
\input{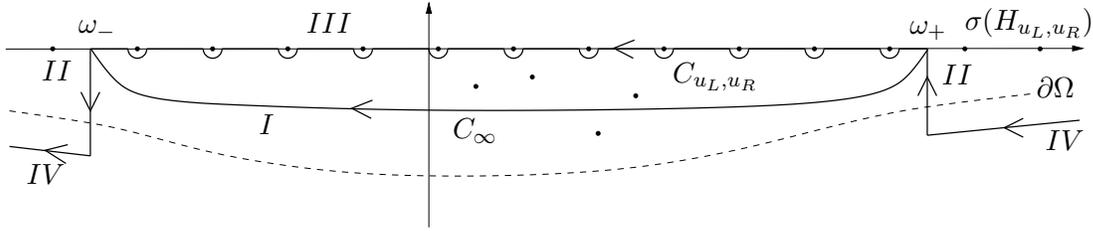}
\caption{Contour deformation onto the real axis}
\label{fig3}
\end{center}
\end{figure}

Using the above analyticity properties of $\bra \Psi, Q_n S_{u_L, u_R} \Psi \ket$ and $\bra \Psi, Q_n S_\infty \Psi \ket$, we are free to deform
the contours $C_{u_L, u_R}$ and $C_\infty$ in any compact set, provided
that $C_{u_L, u_R}$ never intersects $\sigma_n(H_{u_L, u_R})$.
In particular, choosing $\omega_-$ and $\omega_+$ real and outside
of $\sigma(H_{u_L, u_R})$, we may deform the contours as shown in
Figure~\ref{fig3}.
We let $E_{[\omega_-, \omega_+]}$ be the projector on all invariant
subspaces of $H_{u_L, u_R}$ corresponding to real $\omega$ in the range
$\omega_- \leq \omega \leq \omega_+$,
\[ Q_n E_{[\omega_-, \omega_+]} \;=\; \sum_{\omega \in [\omega_-, \omega_+]}
Q_n(\omega)\:E_\omega\:. \]
The next lemma shows that the integral over $C_{III} \cup \overline{C_{III}}$
equals $Q_n E_{[\omega_-, \omega_+]}$, whereas the integrals
over the contours II and IV can be made arbitrarily small by
choosing $|\omega_\pm|$ sufficiently large.
\begin{Lemma} \label{lemma93}
For every $\Psi \in C^\infty_0((u_L, u_R) \times S^2)^2$, $n \in \N$, and
$\varepsilon>0$ there are $\omega_-, \omega_+ \in \R \setminus
\sigma(H_{u_L, u_R})$ such that
\[  \left|
I_n + \bra \Psi, Q_n E_{[\omega_-, \omega_+]} \Psi \ket
\:-\: \frac{1}{2 \pi i} \left(\int_{C_I} \!\!-
\int_{\overline{C_I}} \right) \bra \Psi, Q_n S_\infty \Psi \ket \right| \;\leq\; \varepsilon\:, \]
where $I_n$ are the integrals~(\ref{Indef}) and
$C_I$ is any contour in the lower half plane
which joins $\omega_-$ with $\omega_+$ (see Figure~\ref{fig3}).
\end{Lemma}
{\Proof}
Lemma~\ref{lemma54} yields that by choosing $\omega_+$
and $-\omega_-$ sufficiently large, we can make the contribution of the
contour $IV$ arbitrarily small. The integrals over $C_{III}$ and $\overline{C_{III}}$ combine to contour
integrals around the spectral points on the real axis,
\[ -\frac{1}{2 \pi i} \left( \int_{C_{III}} \!\!- \int_{\overline{C_{III}}} \right) \bra \Psi, Q_n\: S_{u_L, u_R} \Psi \ket\: d\omega
\;=\; -\frac{1}{2\pi i} \sum_{\omega'} \oint_{|\omega-\omega'|=\delta}
\bra \Psi, Q_n\: S_{u_L, u_R} \Psi \ket\: d\omega \:, \]
where the sum runs over all $\omega' \in \sigma(H_{u_L, u_R}) \cap [\omega_-, \omega_+]$, and $\delta$
must be chosen so small that each contour contains only one point of the
spectrum. If we let $\delta \to 0$ and use that $Q_n$ depends smoothly on $\omega$,
one sees that the integrals over the circles converge to
$-2 \pi i \bra \Psi, Q_n\: E_{\omega'} \Psi \ket$. We conclude that
\[ -\frac{1}{2\pi i} \left( \int_{C_{III}} - \int_{\overline{C_{III}}}
\right) \bra \Psi, Q_n\: S_{u_L, u_R} \Psi \ket\: d\omega
\;=\; - \bra \Psi, Q_n\:E_{[\omega_-, \omega_+]} \Psi \ket \,. \]

It remains to show that by choosing $|\omega_\pm|$ sufficiently large, we can
make the integral over the contour $II$ arbitrarily small.
According to Lemma~\ref{lemmaangular}, for sufficiently large $|\omega_\pm|$ the
angular operator ${\cal{A}}_\omega$ is diagonalizable for all~$\omega$ on the contour $II$.
Thus we can assume that the nilpotent matrices ${\cal{N}}$ in the Jordan
decomposition~(\ref{Jordan}) all vanish.
Hence we can separate the resolvents according to Proposition~\ref{thm63} to obtain
\[ \bra \Psi, Q_n\: (S_{u_L, u_R} - S_\infty) \Psi \ket \;=\;
\sum_{\lambda \in \Lambda_n} \bra \Psi, Q_\lambda\:\Delta T_\lambda \Psi \ket \;, \]
where $\Delta T_\lambda$ is the operator with integral kernel
\[ \Delta T_\lambda(u; u', \vartheta') \;=\; (r^2+a^2)^{-\frac{1}{2}}\:
(s_{u_L, u_R} - s_\infty)(u,u') \left( \!\!\begin{array}{cc}
\rho(u', \vartheta') & \sigma(u', \vartheta') \\
\omega\:\rho(u', \vartheta') & \omega\: \sigma(u', \vartheta') \end{array} \!\!\right) . \]
Since the functions $\rho$ and $\sigma$ are smooth and the angular
operators $Q_\lambda$ are bounded (\ref{Qnb}), it suffices to show
that for every $\varepsilon>0$ and $g \in C^\infty_0((u_L, u_R))$, we can
choose $\omega_\pm$ such that for all $\omega$ on the contour $II$,
\[ \int_{-\infty}^\infty du  \int_{-\infty}^\infty du'\: g(u)\: g(u') \:  (s_{[u_L, u_R]}-s_\infty)(u,u') \;\leq\; \varepsilon\:. \]

Let us derive a convenient formula for $s_{[u_L, u_R]}-s_\infty$.
We let $\phi_1$ and $\phi_2$ be the two fundamental solutions which satisfy the
Dirichlet boundary conditions $\phi_1(u_L) = 0 = \phi_2(u_R)$.
Likewise, we let $\acute{\phi}$ and $\grave{\phi}$ be the two fundamental solutions
in infinite volume as constructed in Corollary~\ref{cor74}.
Furthermore, assume that $u_L < u < u' < u_R$. Then, according
to~(\ref{sdef}),
\[ s_{u_L, u_R}(u,u') \;=\; \frac{1}{w(\phi_1, \phi_2)}\: \phi_1(u)\: \phi_2(u') \;,\qquad
s_\infty(u,u') \;=\; \frac{1}{w(\acute{\phi}, \grave{\phi})}\: \acute{\phi}(u)\:
\grave{\phi}(u')\:. \]
Expressing $\phi_1$ as a linear combination of $\acute{\phi}$ and $\phi_2$,
\[ \phi_1(u) \;=\; \acute{\phi}(u)\: \phi_2(u_L) - \acute{\phi}(u_L)\: \phi_2(u)\:, \]
and substituting into the above formula for $s_{u_L, u_R}$, we obtain
\begin{eqnarray*}
s_{u_L, u_R}(u,u') &=& \frac{1}{\phi_2(u_L)\: w(\acute{\phi}, \phi_2)}
\left(\acute{\phi}(u)\: \phi_2(u_L) - \acute{\phi}(u_L)\: \phi_2(u) \right)
\phi_2(u') \\
&=& \frac{1}{w(\acute{\phi}, \phi_2)}\: \acute{\phi}(u)\: \phi_2(u') \:-\:
\frac{\acute{\phi}(u_L)}{\phi_2(u_L)}\: \frac{\phi_2(u)\: \phi_2(u')}{w(\acute{\phi}, \phi_2)} \:.
\end{eqnarray*}
In the first summand, we can express $\phi_2$ in terms of $\acute{\phi}$ and
$\grave{\phi}$,
\beq \label{8rel}
\phi_2(u) \;=\; \acute{\phi}(u)\: \grave{\phi}(u_R) - \acute{\phi}(u_R)\: \grave{\phi}(u) \:.
\eeq
This gives
\begin{eqnarray*}
\frac{1}{w(\acute{\phi}, \phi_2)}\: \acute{\phi}(u)\: \phi_2(u')
&=& \frac{1}{-\acute{\phi}(u_R)\: w(\acute{\phi}, \grave{\phi})} \:
\acute{\phi}(u)
\left( \acute{\phi}(u')\: \grave{\phi}(u_R) - \acute{\phi}(u_R)\: \grave{\phi}(u') \right) \\
&=& -\frac{\grave{\phi}(u_R)}{\acute{\phi}(u_R)}\: \frac{\acute{\phi}(u)\: \acute{\phi}(u')}
{w(\acute{\phi}, \grave{\phi})} \:+\: s_\infty(u,u')\:.
\end{eqnarray*}
We conclude that
\[ (s_{u_L, u_R}-s_\infty)(u,u') \;=\; -\frac{\acute{\phi}(u_L)}{\phi_2(u_L)}\:
\frac{\phi_2(u)\: \phi_2(u')}{w(\acute{\phi}, \phi_2)}
\:-\: \frac{\grave{\phi}(u_R)}{\acute{\phi}(u_R)}\:
\frac{\acute{\phi}(u)\: \acute{\phi}(u')}{w(\acute{\phi}, \grave{\phi})} \:, \]
and because of its symmetry in $u$ and $u'$, this
identity is also valid in the case $u'<u$.
Using~(\ref{8rel}) and the notation~(\ref{7nota}), we get
\beq \label{87a}
(s_{u_L, u_R}-s_\infty)(u,u') \;=\; \frac{\acute{\phi}(u_L)\:
\phi_2(u)\: \phi_2(u')}{\phi_{[u_L, u_R]}\: \acute{\phi}(u_R)\:
w(\acute{\phi}, \grave{\phi})}
\:-\: \frac{\grave{\phi}(u_R)\: \acute{\phi}(u)\: \acute{\phi}(u')}
{\acute{\phi}(u_R)\: w(\acute{\phi}, \grave{\phi})} \:.
\eeq

We choose $|\omega_\pm|$ such that
\[ \int_{u_L}^{u_R} {\mbox{Im}}\, \sqrt{V(\omega_\pm)} \;\in\;
\frac{2 \Z + 1}{4}\: \pi \:. \]
According to the estimate~(\ref{5n2}) in Lemma~\ref{lemma5n}, we can arrange that the
function $\int_{u_L}^{u_R} {\mbox{Im}}\, \sqrt{V}$ is nearly constant on the contour II,
and thus
\beq \label{8z}
\left| \sin \left( 2 \int_{u_L}^{u_R} {\mbox{Im}}\, \sqrt{V} \right) \right| \;\geq\; \frac{1}{2} \:.
\eeq
Propositions~\ref{prp74}, \ref{prp75}, and \ref{prp76} allow us to
estimate each term in~(\ref{87a}) by the corresponding term in the
WKB approximation. According to~(\ref{8z}), the factor
$|\sin (\ldots)|^{-1}$ which appears in Proposition~\ref{prp76} is
bounded. Choosing $\varepsilon$ sufficiently small, We thus obtain the estimate
\begin{eqnarray}
\lefteqn{ \left| (s_{u_L, u_R}-s_\infty)(u,u') \right| } \nonumber \\
&\leq& 2 \left|
\frac{\acute{\alpha}(u_L)\: \alpha_2(u)\: \alpha_2(u')}{\alpha_{[u_L, u_R]}\:
\acute{\alpha}(u_R)\: w(\acute{\alpha}, \grave{\alpha})} \right|\:
\exp \left( 2 \int_{u_L}^{u_R} {\mbox{Re}} \sqrt{V} \right)
+ 2 \left| \frac{\grave{\alpha}(u_R)\: \acute{\alpha}(u)\: \acute{\alpha}(u')}
{\acute{\alpha}(u_R)\: w(\acute{\alpha}, \grave{\alpha})} \right| , \label{8y}
\end{eqnarray}
where we introduced the function
\[ \alpha_2(u) \;=\; |\acute{\alpha}(u)\: \grave{\alpha}(u_R)| +
|\acute{\alpha}(u_R)\: \grave{\alpha}(u)| \:. \]
Using the explicit formulas~(\ref{75x}, \ref{7aLR}) together with~(\ref{8z}), we get
\beq \label{8wex}
|w(\acute{\alpha}, \grave{\alpha})| \;\geq\; |\sqrt{V(u)}|\: |\acute{\alpha}(u)\:
\grave{\alpha}(u)| \:,\qquad
|\alpha_{[u_L, u_R]}| \;\geq\; \frac{1}{2}\: |\acute{\alpha}(u_R)\: \grave{\alpha}(u_R)| \:.
\eeq
Substituting these bounds into~(\ref{8y}), we get an estimate for
$|s_{u_L, u_R}-s_\infty|$ in terms of expressions of the form
\beq \label{712a}
\frac{1}{\sqrt{|V(u)|}}\: \exp \left( 2 \int_{u_L}^{u_R} {\mbox{Re}} \sqrt{V} \right)
\left| \frac{\acute{\alpha}(u_1)}{\acute{\alpha}(v_1)} \right| \cdots \:
\left| \frac{\grave{\alpha}(u_k)}{\grave{\alpha}(v_k)} \right| \cdots
\eeq
with $u_i, v_i \in [u_L, u_R]$. The quotients of the WKB wave functions have
according to~(\ref{81a}) the explicit form
\beq \label{8exp}
\left| \frac{\acute{\alpha}(u)}{\acute{\alpha}(v)} \right| \;=\;
\left| \frac{V(u)}{V(v)} \right|^{-\frac{1}{4}}\:
\exp \left( 2 \int_{u_L}^{u_R} {\mbox{Re}} \sqrt{V} \right) \:,
\eeq
and similarly for $\grave{\alpha}$.
The inequality~(\ref{5n1}) shows that the exponentials in~(\ref{712a})
and~(\ref{8exp}) are bounded uniformly in $\omega$. Furthermore, it is
obvious from~(\ref{5V}) that $V(u)/V(v)$ is close to one if $|\omega|$ is
large. We conclude that on the contour II,
\[ |(s_{u_L, u_R}-s_\infty)(u,u')| \;\leq\; \frac{c}{\sqrt{|V(u)|}} \:, \]
and this can be made arbitrarily small by choosing $|\omega_{\pm}|$
sufficiently large.
\QED

We are now in the position to prove our main theorem. \\[.5em]
{\em{Proof of Theorem~\ref{thm1}. }}
According to Lemma~\ref{lemma93},
\[ -\frac{1}{2 \pi i} \:\lim_{\omega_\pm \to \pm \infty}
\left( \int_{C_I} - \int_{\overline{C_I}} \right)
\bra \Psi, Q_n S_\infty \Psi \ket \:d\omega
\;=\; I_n + \bra \Psi, Q_n E_\R \Psi \ket \;, \]
where $E_\R$ denotes the projector onto the invariant subspace
corresponding to the real spectrum of $H_{u_L, u_R}$.
Here the $\omega_\pm$ are to be chosen as in Lemma~\ref{lemma93} and
$C_I$ is again any contour which joins $\omega_-$ with $\omega_+$
in the lower half plane. Suppose that the contour $C_\varepsilon$
intersects the lines ${\mbox{Re}} \, \omega=\omega_\pm$ in the
points $\omega_+ - i \delta_+$ and $\omega_- - i \delta_-$, respectively.
Then we choose the contour $C_I$ as follows,
\[ C_I \;=\; \left( \omega_- - i [0, \delta_-] \right)
\cup \left( C_\varepsilon \cap (\omega_-, \omega_+) + i \R \right)
\cup \left( \omega_+ - i [0, \delta_+] \right) . \]
The first and last parts of the contour have lengths $\delta_-$ and
$\delta_+$, respectively, and these lengths clearly tend to zero
as $\omega_\pm \to \pm \infty$. Furthermore, it is obvious
from Propositions~\ref{prp74} and~\ref{prp75} as well as~(\ref{8wex}) and~(\ref{8exp}) that the integrand is uniformly bounded on these
parts of the contour. Hence the contribution of these contours tends
to zero as~$\omega_\pm \to \pm \infty$. Thus
\[ -\frac{1}{2\pi i} \left( \int_{C_\varepsilon}-
\int_{\overline{C_\varepsilon}} \right) \bra \Psi,
Q_n S_\infty \Psi \ket \:d\omega \;=\; I_n + \bra \Psi, Q_n E_\R \Psi \ket \;. \]
According to Lemma~\ref{lemma92} and Lemma~\ref{lemmaangular},
the right side of this equation is absolutely summable in $n$ and
\[ \sum_n \left( I_n + \bra \Psi, Q_n E_\R \Psi \ket \right)
\;=\; \bra \Psi, (E_\C + E_\R)\: \Psi \ket \:. \]
Since the spectral projectors in the Pontrjagin space ${\cal{H}}_{u_L, u_R}$
are complete, $E_\C + E_\R=\1$. We conclude that
\[ -\frac{1}{2 \pi i} \sum_n \left( \int_{C_\varepsilon}
-\int_{\overline{C_\varepsilon}} \right) \bra \Psi,
Q_n S_\infty \Psi \ket \:d\omega \;=\; \bra \Psi, \Psi \ket \;. \]
Polarizing, we obtain for every $\Psi \in
C^\infty_0((r_1, \infty) \times S^2)^2$ the simple identity
\begin{equation} \label{sid}
\Psi \;=\; -\frac{1}{2\pi i} \sum_n \left( \int_{C_\varepsilon} -
\int_{\overline{C_\varepsilon}} \right) Q_n S_\infty \Psi\:
d\omega \: .
\end{equation}
The integral and sum converge in $L^2_{\mbox{\scriptsize{loc}}}$.

If we apply the Hamiltonian to the integrand in the above formula,
we obtain according to Proposition~\ref{thm63}
\[ H\: Q_n S_\infty \Psi \;=\; (H-\omega)\: Q_n S_\infty \Psi
\:+\: \omega\: Q_n S_\infty \Psi \;=\; {\mbox{(holomorphic terms)}}
\:+\: \omega\: Q_n S_\infty \Psi \:. \]
The holomorphic terms are holomorphic in the whole neighborhood
of the real axis
enclosed by $C_\varepsilon$ and $\overline{C_\varepsilon}$
(see Lemma~\ref{lemmaangular} {\bf{(i)}}),
and therefore the contour integral over them drops out.
We conclude that applying $H$ reduces to multiplying the
integrand by a factor $\omega$. Iteration shows that the
dynamics of $\Psi$ is taken into account by a factor
$e^{-i \omega t}$,
\[ \Psi(t) \;=\; -\frac{1}{2\pi i} \sum_n \left( \int_{C_\varepsilon} -
\int_{\overline{C_\varepsilon}} \right) e^{-i \omega t}\:
Q_n S_\infty \Psi_0 \: d\omega \: . \]
Comparing this expansion with~(\ref{sid}), one sees that the integrand in the
last expansion is equal to the integrand in~(\ref{sid}) if
$\Psi$ is replaced by $\Psi(t)$.
Since $\Psi(t)$ is smooth and by causality has compact support,
we conclude that the integral and sum again converge in $L^2_{\mbox{\scriptsize{loc}}}$.
Finally, using that
the contour integrals in this formula are all independent of
$\varepsilon$, we may take the limit $\varepsilon \searrow 0$ of each of them.
\QED

\noindent
{\em{Acknowledgments:}} We are grateful to Harald Schmid and Johann
Kronthaler for helpful discussions. We would like to thank McGill University,
Montr{\'e}al, the Max Planck Institute for Mathematics in the Sciences, Leipzig, and the University of Michigan for support and hospitality. We are also grateful to the Vielberth Foundation, Regensburg, for generous support.

\begin{tabular}{lcl}
\\
Felix Finster & $\;\;\;\;$ & Niky Kamran\\
NWF I -- Mathematik && Department of Math.\ and Statistics \\
Universit{{\"a}}t Regensburg && McGill University \\
93040 Regensburg, Germany && Montr{\'e}al, Qu{\'e}bec \\
{\tt{Felix.Finster@mathematik}} && Canada H3A 2K6  \\
$\;\;\;\;\;\;\;\;\;\;\;\;\;\;$ {\tt{.uni-regensburg.de}}
&& {\tt{nkamran@math.McGill.CA}} \\
\\
Joel Smoller & $\;\;$ & Shing-Tung Yau \\
Mathematics Department && Mathematics Department \\
The University of Michigan && Harvard University \\
Ann Arbor, MI 48109, USA && Cambridge, MA 02138, USA \\
{\tt{smoller@umich.edu}} && {\tt{yau@math.harvard.edu}}
\end{tabular}

\end{document}